\documentclass[aps,prb,showpacs,twoside,twocolumn]{revtex4}
\usepackage{graphicx}

\begin{document}

\title{Aharonov-Bohm ring with fluctuating flux}

\author{Florian Marquardt and C. Bruder}
\affiliation{Department of Physics and Astronomy, University of Basel,
 Klingelbergstrasse 82, CH-4056 Basel, Switzerland}

\date{November 8th, 2001}

\begin{abstract}
We consider a non-interacting system of electrons on a clean one-channel Aharonov-Bohm
ring which is threaded by a fluctuating magnetic flux. The flux derives from
a Caldeira-Leggett bath of harmonic oscillators. We address the influence of
the bath on the following properties: one- and two-particle Green's functions,
dephasing, persistent current and visibility of the Aharonov-Bohm effect in
cotunneling transport through the ring. For the bath spectra considered here
(including Nyquist noise of an external coil), we find no dephasing in the \emph{linear}
transport regime at \emph{zero} temperature.
\end{abstract}


\maketitle

\section{Introduction}

\begin{figure}
\centerline{\hbox{  \includegraphics[width=6cm]{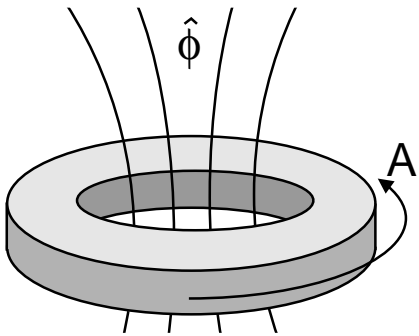}}}

\caption{\label{Abb:model}The model situation: A fluctuating flux leads, via a time-dependent
vector potential, to a fluctuating force for the electrons on the Aharonov-Bohm
ring}
\end{figure}

In the present work, we consider a simple theoretical model system of non-interacting spinless
electrons which are restricted to move in one dimension around the circumference
of a clean, one-channel Aharonov-Bohm ring. The ring is threaded by a magnetic
flux which fluctuates around some average value (see Fig.~\ref{Abb:model}).
This may lead to dephasing of the electron motion on the ring, apart from other
effects like renormalization of the electron masses and introduction of an effective
coupling between the electrons. We treat the full dynamics of the fluctuating
flux coupled to the electron system in a selfconsistent manner, rather than
prescribing an external stochastic time-dependent classical field. In order
to achieve this, the flux is taken to be the sum of the normal coordinates of
a Caldeira-Leggett type bath of harmonic oscillators \cite{2,weiss}.
The fluctuations couple to the electrons via the vector-potential term in the
kinetic energy. As an important special case for the bath spectrum we treat
the Nyquist noise that may be due to the equilibrium current fluctuations in
the external coil producing the flux.

Equilibrium and transport properties of this model system are analyzed for the
cases of zero and finite temperatures, taking into account the coupling to the
bath and the Pauli principle with respect to the electron system. In particular,
we discuss the single-particle and two-particle Green's functions, level widths,
energy shifts and dephasing times, and the reduction of the persistent current
due to the fluctuations. In each case, the dependence on the coupling strength
between system and bath and on the low-frequency spectral properties of the
bath is examined. Aharonov-Bohm interference observed in cotunneling through
the ring is discussed in order to analyze the coherence properties of the electrons
on the ring under the influence of the fluctuating flux in a transport situation.
As a result of our calculation, we find that the fluctuations do \textit{not}
lead to dephasing in the \textit{linear} transport regime at \textit{zero} temperature.

The single-particle version of this model has been considered before in Ref.
\onlinecite{park} in order to determine whether persistent currents in a normal metal
ring may be destroyed by coupling to an Ohmic bath. Whereas the authors of Ref.~\onlinecite{park}
used the Feynman-Vernon influence functional\cite{1}, we will apply a different, more
direct approach. We emphasize that our analysis is restricted to baths
weaker than the Ohmic bath (see the discussion after Eq. (\ref{ratio})). The possibility of \emph{spontaneous} persistent currents was
investigated (and ruled out) in Ref.~\onlinecite{lossmartin}, using a Luttinger liquid
picture for the electrons and taking into account their electromagnetic self-interaction.
 Dephasing of a single electron going around the two arms
of an Aharonov-Bohm ring has been considered both in Ref.~\onlinecite{lossmullen}, using
the influence functional, and in Ref.~\onlinecite{sai}, using a semiclassical picture.
In the latter article, the connection between phase fluctuations and the trace
left by the system in the environment was emphasized. More recently, the question
of dephasing in mesoscopic systems has received renewed attention due to 
a set of weak-localization measurements which have shown a saturation of the dephasing
time at low temperatures \cite{mohanty,cohenimry}. Motivated in part by these puzzling findings,
the authors of Refs. \onlinecite{cedraschi,buettrevb,buettnato} considered a ring containing
a single quantum dot with fluctuating gate voltage and obtained the properties
of the quantum-mechanical ground state (in particular the persistent current).
A strong influence of external \emph{nonequilibrium} noise on the persistent
current in a \emph{disordered} \emph{quasi}-1D ring has been found recently
in Ref.~\onlinecite{kravtsov}. The effects of a phase-breaking scatterer in the many-particle situation, where the Pauli principle becomes important, have been discussed in Ref.~\onlinecite{mello}. Very recently, dephasing in a mesoscopic Mach-Zehnder type
interference setup has been analyzed in Ref.~\onlinecite{buettseelig}.

In the following section, we will define the model and discuss some simple consequences
as well as some features which cannot be included in this system easily. Then
we give a short qualitative discussion of dephasing for the simplified case
of a classical fluctuating flux (represented by a random process). Similar considerations
are applied to the calculation of the Green's function for a single electron
on the ring, both with classical and quantum fluctuations. The resulting energy
shifts and level shapes are analyzed in some detail, since these results can
be taken over to the many-particle calculation of the single-particle and two-particle
Green's functions which is presented in section \ref{Sec:manyparticle} together
with the evaluation of the grand canonical partition sum and the persistent
current. After discussing the physical meaning of the dephasing produced by
the Nyquist noise at low temperatures, we turn to an analysis of Aharonov-Bohm
interference in a cotunneling transport measurement.

\section{The model}
\label{modelsec}

The Hamiltonian of the system of electrons on the ring is given by

\begin{equation}
\label{thehamiltonian}
\hat{H}\equiv \sum _{p}\hat{\Psi }^{\dagger }_{p}\frac{\left( p-g\hat{\phi }\right) ^{2}}{2m}\hat{\Psi }_{p}+\hat{H}_{bath}\, ,
\end{equation}

where \( \hat{H}_{bath} \) is the Hamiltonian of the set of uncoupled oscillators
representing the bath:

\begin{equation}
\label{hambath}
\hat{H}_{bath}\equiv \sum ^{N_{osc}}_{j=1}\left\{ \frac{\hat{P}^{2}_{j}}{2M}+\frac{M\omega _{j}^{2}}{2}\hat{Q}_{j}^{2}\right\} \, .
\end{equation}

The possible values of the electron momentum \( p \) are quantized due to the
finite circumference \( L \) of the ring: \( p=2\pi n/L \) with an integer
\( n \). (Note that here and in the following, we will put \( \hbar \equiv k_{B}\equiv 1 \))
The term \( g\hat{\phi } \) in the kinetic energy of the electrons is due to
the coupling to the vector potential which is proportional to the fluctuating
flux. \( \hat{\phi } \) represents this flux (up to a constant factor) and
is assumed to be given by the sum over the oscillator normal coordinates:

\begin{equation}
\label{testmark}
\hat{\phi }\equiv \frac{1}{\sqrt{N_{osc}}}\sum _{j}\hat{Q}_{j}\, .
\end{equation}

The prefactor in this definition has been chosen such that the autocorrelation
function \( \left\langle \hat{\phi }(t)\hat{\phi }(0)\right\rangle  \) of \( \hat{\phi } \)
has a well-defined limit if the number \( N_{osc} \) of oscillators tends to
infinity while the spacing of frequencies tends to zero like \( 1/N_{osc} \).
This is the {}``thermodynamic limit{}'' of an infinite bath which is necessary
to describe truly irreversible, dissipative dynamics. The quantity \( g \)
is the coupling strength between bath and electrons. It incorporates the electron
charge and the circumference of the ring, since the line-integral of the vector
potential around the ring gives the flux. Any external \emph{static} magnetic
flux \( \Phi  \) has to be added in the kinetic energy expression. 

We assume the interaction between system and bath to be sufficiently
weak, such that the bath may be treated as linear in a good approximation 
(like it is usually
done in the theory of quantum dissipative systems, see Ref. \onlinecite{2}
for a more detailed discussion).
Apart from this assumption, the expression
Eq.~(\ref{testmark}) used for the fluctuating flux is still
completely general. We are free to choose the frequencies of the
bath oscillators to obtain any desired
correlation function of $\hat{\phi}$, which is the only quantity that
affects the dissipative system dynamics in this model. Note as well
that the coupling of the velocity to a vector potential assumed in
Eq.~(\ref{thehamiltonian}) has been shown\cite{Ford} to be equivalent
to a Caldeira-Leggett model with the usual coordinate-type coupling 
for a particle moving on an infinite line. Physically, the 
coupling used here is the natural choice for a situation in which the vector
potential is linearly related to the fluctuating current in an
external coil. The fluctuations (and linear response) of the current
then determine the bath correlator discussed in the next
section.

The correlator of \( \hat{\phi } \) will determine the dephasing rate and other
important quantities via its low-frequency properties. For the discrete set
of oscillators and for the continuum limit, respectively, it is given by the
following expressions, where the average \( \left\langle \cdot \right\rangle  \)
is taken with respect to the unperturbed set of oscillators:

\begin{eqnarray}
 &  & \left\langle \hat{\phi }(t)\hat{\phi }(0)\right\rangle \nonumber \\
 & = & \frac{1}{N_{osc}}\sum _{j}\frac{1}{2M\omega _{j}}\left[ \coth \left( \frac{\omega _{j}}{2T}\right) \, \cos \left( \omega _{j}t\right) -i\sin \left( \omega _{j}t\right) \right] \nonumber \\
 & = & \int _{0}^{\infty }d\omega \, C(\omega )\, \left[ \coth \left( \frac{\omega }{2T}\right) \, \cos \left( \omega t\right) -i\sin \left( \omega t\right) \right] \, .\label{phicorrmark} 
\end{eqnarray}

This defines the spectral function \( C(\omega ) \) which we will use to characterize
the bath spectrum. In terms of the discrete set of frequencies, it is given
by \( C(\omega )\equiv N_{osc}^{-1}\cdot \sum _{j}\delta \left( \omega -\omega _{j}\right) \left( 2M\omega _{j}\right) ^{-1} \).
Note that the \( \coth  \) is equal to \( 2n(\omega )+1 \), where \( n(\omega ) \)
is the Bose-Einstein distribution function. The \emph{special case of Nyquist
noise} is obtained by the requirement that for high temperatures, the spectrum
of fluctuations of the magnetic flux (i.e. of \( \hat{\phi } \)) is white.
Since \( \coth \left( \omega /2T\right) =2T/\omega  \) for \( T\gg \omega  \),
this means \( C(\omega )\propto \omega  \) (for small \( \omega  \)).

Before we proceed to the calculations, we will point out some simplifying features
of this situation as well as some important aspects of the dephasing problem
in degenerate Fermion systems that are beyond the scope of this model.

The magnetic flux is assumed to thread the ring in such a way that the situation
is axially symmetric with respect to the axis which goes through the center
of the ring and is perpendicular to its plane. In this case, we can choose the
gauge such that the vector potential is everywhere tangential to the ring and
of constant magnitude around the whole circumference. The same holds for the
electric field, which is given by the time-derivative of the vector potential.
This is analogous to the Caldeira-Leggett treatment of one-dimensional quantum
Brownian motion of a free particle, with the formal difference that in our case
the force is derived from a vector potential instead of a scalar potential\cite{park,Ford}.
It is the choice appropriate for a system with periodic boundary conditions,
where the quantization of momenta, the Aharonov-Bohm effect and persistent currents
play a role. Note that under different circumstances the assumption of a force
which is constant in space is only valid within the dipole approximation for
a particle that is restricted to move in a well-localized region of space, because
otherwise the finite wavelength of the bath modes (phonons etc.) becomes important.

Since the coupling between system and bath is via the momentum, which commutes with the electron Hamiltonian and therefore 
is a constant of the motion for the original uncoupled electron system ({}``diagonal
coupling{}''), some features of this model are very simple: In spite of the
interaction with the bath, the momentum of a particle will stay constant, only
its velocity and kinetic energy can fluctuate. This simplifies the many-electron
problem as well: Although the fluctuating force influences the center-of-mass
motion of the electrons and introduces some kind of {}``effective interaction{}''
between them, the occupation of different \( p \)-states cannot be changed
by the bath. Note that this simplification would be spoiled if one takes into
account impurities and/or a coupling that depends on the position. For example,
the latter would arise if one considered an arbitrary fluctuating electromagnetic
field or the electric field between the plates of a capacitor, which is a constant
vector field in space but is not constant with respect to its projection onto
the direction of motion of the electrons on the ring. Other situations where
the coupling depends on position include interaction with phonons or a localized
spin on the ring.

In its single-particle version, the Hamiltonian given above also arises in the
discussion of dephasing for a charged island, if the coupling to the bath (e.g.
a fluctuating gate voltage) is purely diagonal in the system's eigenbasis and
can only lead to fluctuations in the energy levels. Note, however, that questions
of interest here like tunneling into the ring or features of the many-particle
system have no natural counterpart in that rather simple situation.

Although in our model all electrons are coupled to the same flux, which introduces
a kind of effective interaction between them, the decay rates of Green's functions
will not show any dependence on the distance to the Fermi surface, in contrast
to the usual behaviour of interacting Fermi systems. This is due to the diagonal
coupling between system and bath, which means that there are no energy-relaxation
processes which change the occupation numbers of the electrons and which would
feel the restriction by the Pauli principle. A related question arises in the
study of dephasing in degenerate Fermion systems: If the coupling is not diagonal
in the electrons' (single-particle) eigenstates, the system variable which couples
to the bath (in our case the momentum) carries out fluctuations itself. In a
semiclassical single-particle calculation, these fluctuations pick out the high-frequency
components of the bath spectrum\cite{cohenimry}. Therefore, according to such a calculation,
there is dephasing even in the case of a bath spectrum that vanishes at low
frequencies and this implies that at low temperatures, the high-frequency zero
point fluctuations of the bath contribute heavily to dephasing \emph{in this
picture}. However, if the electron system is nearly degenerate, many of its
transitions will be blocked by the Pauli principle so that such an effect will
be strongly suppressed. Although we cannot investigate this point in our model,
related considerations will occur in our discussion of the cotunneling setup
in section \ref{abcotunnel}.

\section{Single-particle problem}

In this section we will discuss the problem of a single particle on the ring,
both semiclassically and quantum-mechanically. The results can support the understanding
of the next section, which is devoted to the many-particle situation.

\subsection{Semiclassical analysis}

Consider two wave packets traversing the left and right arm of the Aharonov-Bohm
ring with constant velocity \( v \) and meeting again after some time \( t=L/\left( 2v\right)  \)
at the opposite end. The resulting interference pattern depends on the total
phase difference between the two paths. In a semiclassical calculation, the
phase difference is produced by the \( vA \)-term in the Lagrangian of the
particle, and it is given by

\begin{equation}
\varphi =2\frac{e}{c}\int _{0}^{t}v\, A(t')dt'\, .
\end{equation}

The factor \( 2 \) arises because the phases are equal up to a change in sign.
\( A(t') \) gives the time-dependence of the fluctuating vector potential which
is assumed to be a classical Gaussian random process with zero mean in the high-temperature limit considered here. The visibility
of the interference pattern will be suppressed due to the fluctuations in the
phase \( \varphi  \). Since \( \varphi  \) is a Gaussian random variable,
we obtain for the suppression factor

\begin{equation}
\label{gaussianaverage}
\left\langle e^{i\varphi }\right\rangle =e^{-\left\langle \varphi ^{2}\right\rangle /2}\, .
\end{equation}

In our model \( eA/c \) is equal to \( g\phi  \), where we will treat \( \phi  \)
as a classical fluctuating field by taking into account only the real (symmetric)
part of the correlator (\ref{phicorrmark}) in the high-temperature limit. Then
the variance \( \left\langle \varphi ^{2}\right\rangle  \) of the phase becomes:

\begin{equation}
\left\langle \varphi ^{2}\right\rangle =4g^{2}v^{2}\int _{0}^{t}dt_{1}\int _{0}^{t}dt_{2}\, \left\langle \phi (t_{1})\phi (t_{2})\right\rangle \, .
\end{equation}

If the traversal time \( t \) is much larger than the correlation time of the
fluctuations in \( \phi  \), we may apply the following standard approximation:

\begin{eqnarray}
\left\langle \varphi ^{2}\right\rangle  & \approx  & t\cdot 4g^{2}v^{2}\int _{-\infty }^{+\infty }dt'\, \left\langle \phi (t')\phi (0)\right\rangle \nonumber \\
 & = & t\cdot 4g^{2}v^{2}\cdot 4\pi T\left( \frac{C(\omega )}{\omega }\right) _{\omega \rightarrow 0}\, .\label{phaseuncertaintyclassical} 
\end{eqnarray}

This means that in the case of Nyquist noise (\( C\propto \omega  \)) we obtain
a finite {}``dephasing rate{}'' that grows linearly with temperature. Note
however, that the time \( t \) introduced here cannot grow without bounds but
is fixed by the circumference \( L \) and the velocity \( v \). This calculation
already shows that the dephasing rate will vanish for \( v\rightarrow 0 \)
or \( T\rightarrow 0 \). For a bath that is weaker at low frequencies (\( C\propto \omega ^{\alpha } \)
with \( \alpha >1 \)) we do not obtain a suppression factor that decays exponentially
with time, hence the dephasing rate is always zero. We defer a more detailed
discussion of the various cases to the full quantum-mechanical treatment below. 

The physical interpretation of this result is clear: The fluctuating electric
field \( \propto \dot{A} \) leads to a fluctuating velocity \( \propto A \),
so that the random shift in the interference pattern is \( \Delta x\propto \int A(t')\, dt' \).
The interference pattern will be completely washed out once the spread in \( \Delta x \)
becomes comparable to the wavelength \( \lambda \propto 1/v \). This coincides
with the criterion \( \left\langle \varphi ^{2}\right\rangle \approx 1 \).

Very similar considerations arise in the calculation of the Green's function
of a single electron on the ring. The retarded Green's function may be approximated
semiclassically by averaging the amplitude for propagation of the electron under
the influence of the fluctuating flux:

\begin{eqnarray}
iG_{p}^{R}(t) & = & \theta (t)\left\langle \left\{ \hat{\Psi }_{p}(t),\hat{\Psi }_{p}^{\dagger }(0)\right\} \right\rangle \nonumber \\
 & \approx  & \left\langle \exp \left[ -i\int ^{t}_{0}\frac{\left( p-g\phi (t')\right) ^{2}}{2m}dt'\right] \right\rangle \, .
\end{eqnarray}

The exponential contains a quadratic term \( \left( g\phi \right) ^{2} \) which
does not represent a Gaussian random variable, so that formula (\ref{gaussianaverage})
cannot be applied to perform the averaging. Here and in the following quantum-mechanical
calculation, we will neglect this term, which does not couple to the momentum
(cf. discussion in section \ref{quadraticfluxterm}). With this approximation,
the Green's function is given by an expression which again involves the correlation
function of \( \phi  \):

\begin{eqnarray}
& & iG^{R}_{p}(t)\approx \exp \left[ -i\frac{p^{2}}{2m}t \right] \nonumber \\
& &
\times \exp \left[-\frac{1}{2}\left( \frac{gp}{m}\right) ^{2}\int _{0}^{t}dt_{1}\int _{0}^{t}dt_{2}\, \left\langle \phi (t_{1})\phi (t_{2})\right\rangle \right] \, . \label{formula10}
\end{eqnarray}

At this point, the discussion given above applies. In particular, for Nyquist
noise and finite temperatures, the Green's function decays exponentially with
a rate proportional to \( g^{2}p^{2}T \).

\subsection{Single-particle Green's function: Quantum case}

In the following, we will calculate and discuss the quantum-mechanical Green's
function of a \emph{single} electron on the ring. The extension to the many-particle
case will be given in the next section. The single-particle density of states (DOS),
which is given by the imaginary part of the Fourier transform of \( G^{R} \),
is a measurable quantity, as it can be revealed by tunneling into the ring.

We imagine a situation where the ring is empty and a single electron is inserted,
so that the retarded Green's function is

\begin{equation}
iG^{R}_{p}(t)=\theta (t)\left\langle \hat{\Psi }_{pt}^{}\hat{\Psi }^{\dagger }_{p0}\right\rangle =\theta (t)\left\langle e^{i\hat{H}t}\hat{\Psi }_{p}e^{-i\hat{H}t}\hat{\Psi }_{p}^{\dagger }\right\rangle \, .
\end{equation}

The average is a thermal expectation value with respect to the unperturbed bath
of oscillators, corresponding to the situation without any particle on the ring.
Obviously the bath cannot change the occupation of the different momentum states.
Therefore, we only have to take into account that the time evolution between
creation and destruction of the electron is governed by the Hamiltonian which
contains the kinetic energy term that couples to the bath via the flux \( \hat{\phi } \).
Through this coupling, the introduction of the particle into the ring perturbs
the bath oscillators. In anticipation of the many-electron case, we will denote
by \( \hat{H}[\{p_{j}\}] \) the Hamiltonian for a fixed number of occupied
momentum states, given by the set \( \{p_{j}\} \). This operator only acts
on the bath. Then the matrix element of the time-evolution operator with respect
to the electron system is given by the following expression, for a Slater determinant
belonging to the configuration \( \left\{ p_{j}\right\}  \):

\begin{equation}
\left\langle \{p_{j}\}\right| e^{-i\hat{H}t}\left| \{p_{j}\}\right\rangle =e^{-i\hat{H}[\{p_{j}\}]t}\, .
\end{equation}
 In particular, without any electrons we have \( \hat{H}[\emptyset ]\equiv \hat{H}_{bath} \).
For the Green's function considered here, this leads to:

\begin{eqnarray}
\left\langle \hat{\Psi }_{pt}^{}\hat{\Psi }^{\dagger }_{p0}\right\rangle  & = & \left\langle e^{i\hat{H}[\emptyset ]t}e^{-i\hat{H}[\{p\}]t}\right\rangle \nonumber \\
 & = & \left\langle \hat{T}\, \exp \left[ -i\frac{p^{2}}{2m}t+i\frac{gp}{m}\int _{0}^{t}dt'\, \hat{\phi }(t')\right] \right\rangle \, .
\end{eqnarray}
 This expectation value can be interpreted as the (thermally averaged) overlap
between the initial bath state evolved once with and once without presence of
a particle on the ring. In the second line, we have introduced the interaction
picture with respect to \( \hat{H}_{bath} \) and dropped the term quadratic
in \( g\hat{\phi } \) from the kinetic energy (compare the discussion above).
Since \( \hat{\phi } \) is a bosonic variable (i.e. linear in the oscillator
normal coordinates), Wick's theorem can be applied to the evaluation of this
time-ordered thermal average, using a linked-cluster expansion. It leads to
an expression completely analogous to the one used above for the classical Gaussian
random process, see (\ref{gaussianaverage}) and (\ref{formula10}). The difference
consists in the replacement of the classical correlator by the thermal time-ordered
expectation value:

\begin{eqnarray}
 & \left\langle \hat{T}\, \exp \left[ i\kappa \int _{0}^{t}dt'\, \hat{\phi }(t')\right] \right\rangle  & \nonumber \\
= & \exp \left[ -\frac{\kappa ^{2}}{2}\int _{0}^{t}dt_{1}\int _{0}^{t}dt_{2}\, \left\langle \hat{T}\hat{\phi }(t_{1})\hat{\phi }(t_{2})\right\rangle \right] \, . & \label{timeorder} 
\end{eqnarray}

At present, \( \kappa =gp/m \), but the same formula will be used below with
other values for \( \kappa  \). In contrast to the classical correlator, \( \left\langle \hat{T}\hat{\phi }(t_{1})\hat{\phi }(t_{2})\right\rangle  \)
is complex and will lead to an energy shift in addition to a decay of the Green's
function. Using

\begin{eqnarray}
\left\langle \hat{T}\hat{\phi }(t)\hat{\phi }(0)\right\rangle = &  & \nonumber \\
\int _{0}^{\infty }d\omega \, C(\omega )\, \left[ \left( 2n(\omega )+1\right) \cos \left( \omega t\right) -i\sin \left( \omega \left| t\right| \right) \right]  &  & 
\end{eqnarray}

to evaluate the double time-integral, we obtain for the exponent of Eq. (\ref{timeorder}):

\begin{eqnarray}
i\, t\kappa ^{2}\int _{0}^{\infty }d\omega \, \frac{C(\omega )}{\omega } &  & \nonumber \\
+\kappa ^{2}\int _{-\infty }^{+\infty }d\omega \, \frac{C(\left| \omega \right| )}{\omega ^{2}}\left( n\left( \left| \omega \right| \right) +\theta (\omega )\right) \left( e^{-i\omega t}-1\right) \, . &  & \label{greenexponent} 
\end{eqnarray}

The step-function \( \theta (\omega ) \) corresponds to the zero-point fluctuations.
In the remainder of this section, we will discuss the behaviour of the Green's
function derived from (\ref{greenexponent}) at short and long times, for different
bath spectra.

At short times, the integral in the second line compensates the first integral,
so that the exponent begins to grow like \( t^{2} \) instead of \( t \). Physically,
this means that at short times the particle has not yet influenced the bath
and its energy is still given by the bare energy \( p^{2}/2m \). At later times,
the bath oscillators have been shifted by the presence of the particle, so the
overlap between initial and final bath state is diminished and the Green's function
decays. The first integral produces a negative energy shift, corresponding to
the formation of a new interacting state of particle and bath. Since this shift
is proportional to \( p^{2} \), it is equivalent to an enhanced effective mass
\( m^{*} \). 

Combining the energy-shift with the initial kinetic energy \( p^{2}/2m \),
we obtain the following expression for \( m^{*} \):

\begin{equation}
\label{effectivemass}
\frac{1}{m^{*}}=\frac{1}{m}\left( 1-2\frac{g^{2}}{m}\int _{0}^{\infty }d\omega \, \frac{C(\omega )}{\omega }\right) \equiv \frac{1}{m}\left( 1-\xi \right) \, .
\end{equation}

Alternatively, the effective mass can be calculated from the initial Hamiltonian
(for a single particle) by using the term \( gp\hat{\phi }/m \) from the particle's
kinetic energy to introduce a momentum-dependent shift into the oscillator potential
energies of Eq. (\ref{hambath}): \( \hat{Q}_{j}\mapsto \hat{Q}_{j}-gp(mM\omega _{j}^{2}\sqrt{N_{osc}})^{-1} \).
The resulting term quadratic in \( p \) must be compensated for, which yields
the change in mass:

\begin{equation}
\label{effectivemass2}
\frac{1}{m^{*}}=\frac{1}{m}\left( 1-\frac{g^{2}}{mM}\frac{1}{N_{osc}}\sum \frac{1}{\omega _{j}^{2}}\right) \, .
\end{equation}

Since we have neglected the term \( \left( g\hat{\phi }\right) ^{2} \) in the
preceding derivation, the effective mass displayed here is only correct in lowest
order with respect to \( g^{2} \). A full calculation yields \( m^{*}=m\left( 1+\xi \right)  \).

The real part of the second integral in Eq. (\ref{greenexponent}) gives a negative
contribution, corresponding to a suppression in magnitude of the Green's function.
If the bath is relatively weak at low frequencies (\( C\propto \omega ^{\alpha } \)
with \( \alpha \geq 3 \)), the long-time behaviour is simple: The decay saturates
at \( t\rightarrow \infty  \).

\begin{figure}
\centerline{\hbox{\includegraphics[width=4cm]{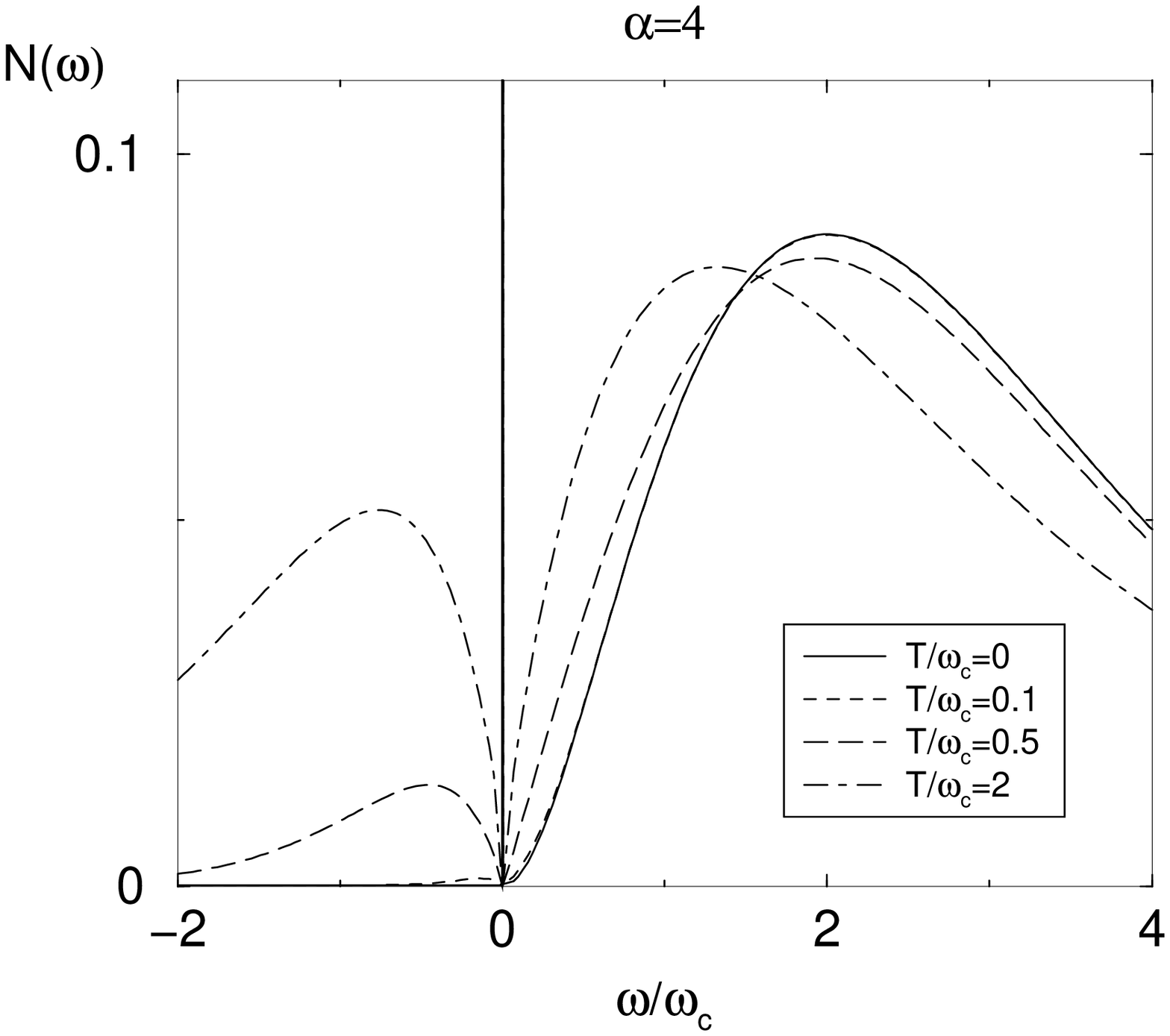} \, \includegraphics[width=4cm]{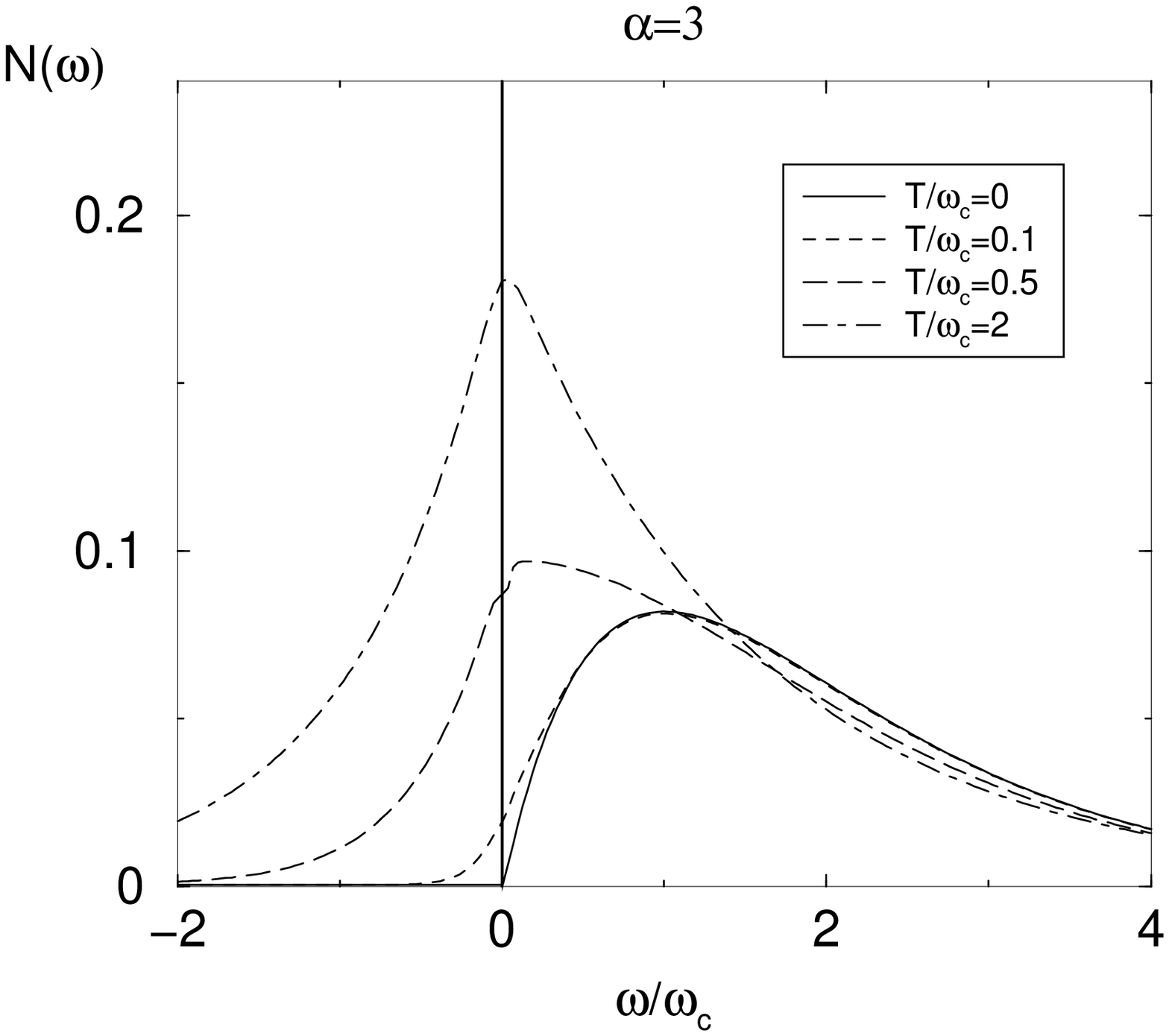} }}
\centerline{\hbox{\includegraphics[width=4cm]{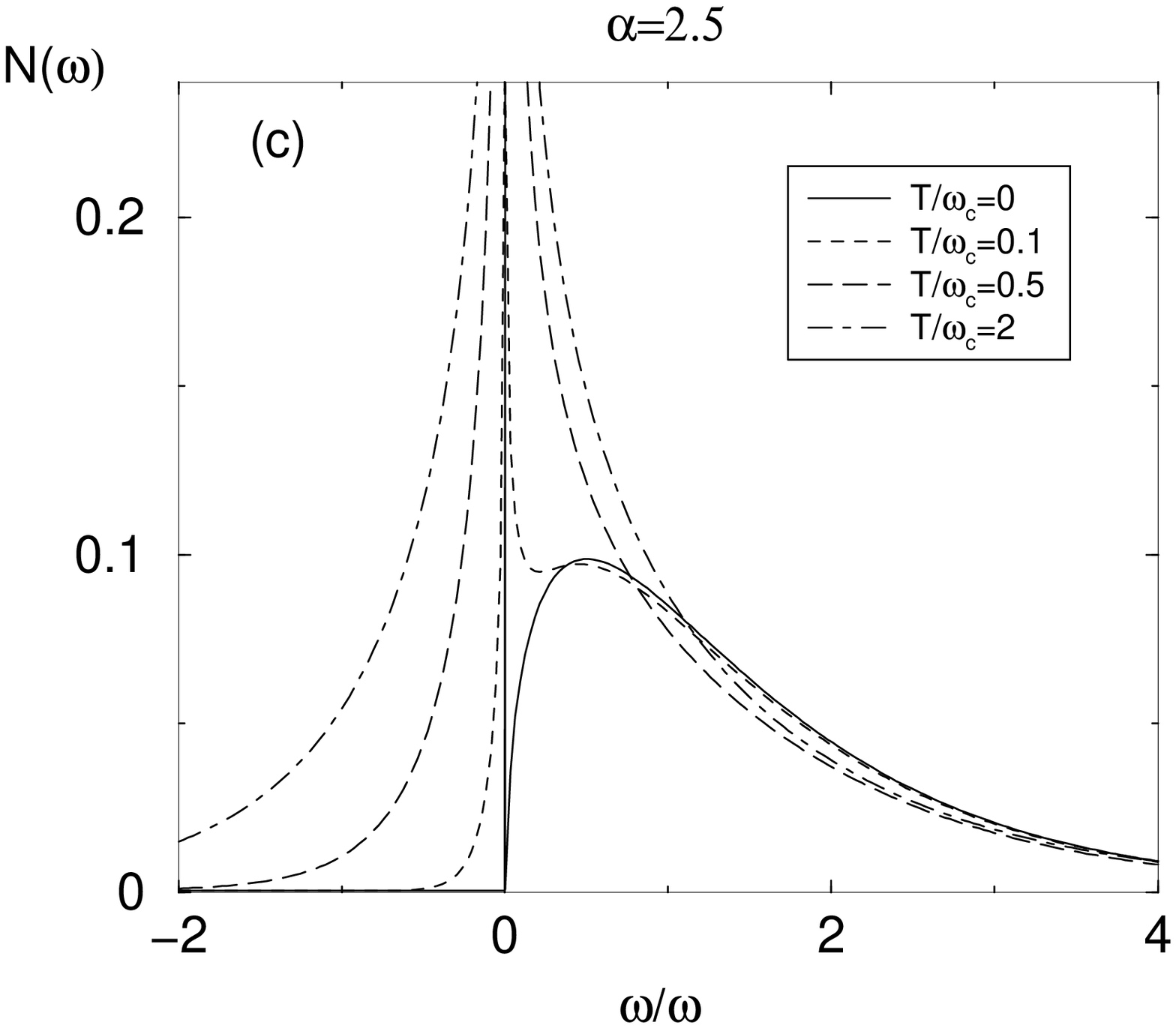} \, \includegraphics[width=4cm]{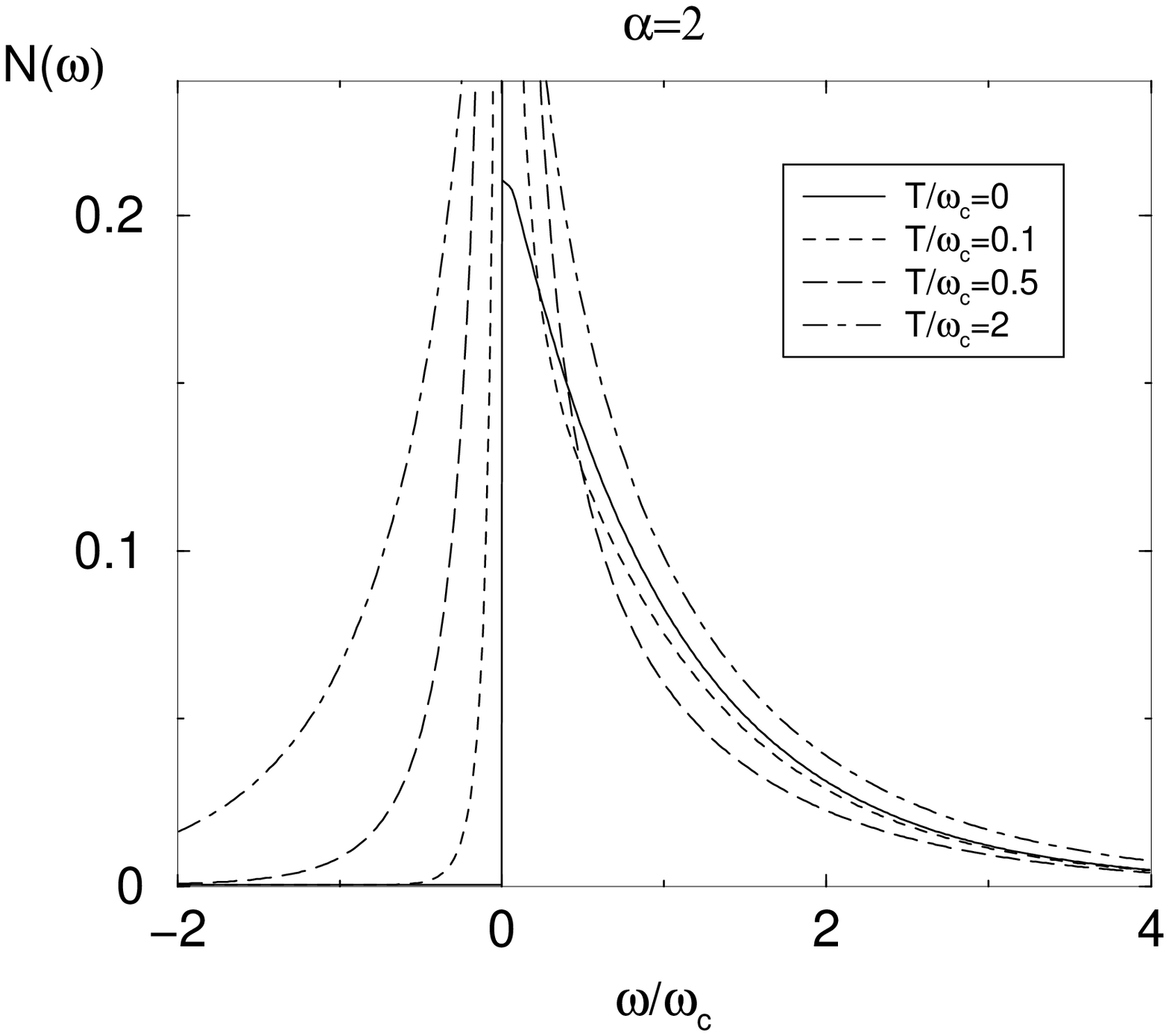} }}

\caption{\label{Abb:dos}The density of states for different bath spectra and temperatures. The quantity \protect\( N(\omega )\equiv \Im mK(\omega )/\pi \protect \),
which corresponds to the single-particle DOS obtained from the retarded Green's
function, is plotted vs. frequency \protect\( \omega /\omega _{c}\protect \),
for different exponents \protect\( \alpha \protect \) of the bath spectrum
at low frequencies (\protect\( C(\omega )\propto \omega ^{\alpha } \exp(-\omega/\omega_c )\protect \))
and for different temperatures. At \protect\( T=0\protect \), the DOS goes
like \protect\( \omega ^{\alpha -2}\protect \) at low \protect\( \omega \protect \)
and vanishes for \protect\( \omega <0\protect \). In all cases displayed here,
there is a delta peak at \protect\( \omega =0\protect \), except for \protect\( \alpha =2\protect \)
at \protect\( T>0\protect \). In the limit \protect\( T\rightarrow \infty \protect \)
the level shape becomes symmetric in all cases and the strength of the delta
peak vanishes. }
\end{figure}

In this case, the Fourier transform of the Green's function,

\begin{equation}
G^{R}_{p}(\omega )=\int _{0}^{\infty }dt\, e^{i\omega t}G^{R}_{p}(t)\, ,
\end{equation}

still has a {}``quasiparticle{}'' delta peak in the density of states \( \Im m\, G^{R}_{p}(\omega )/\pi  \),
but of a reduced magnitude. It is superimposed on an {}``incoherent background{}''
(see Fig.~\ref{Abb:dos}).

For later use, we define the Fourier transform of the Green's function (without
the constant energy shift):

{\footnotesize \begin{eqnarray}
iK(\omega ;\kappa )\equiv \int _{0}^{\infty }dt\, e^{i\omega t}\,  &  & \nonumber \\
\times \exp \left[ \kappa ^{2}\int _{-\infty }^{+\infty }d\tilde{\omega }\, \frac{C(\left| \tilde{\omega }\right| )}{\tilde{\omega }^{2}}\left( n\left( \left| \tilde{\omega }\right| \right) +\theta (\tilde{\omega })\right) \left( e^{-i\tilde{\omega }t}-1\right) \right] \, . &  & \label{Kdef} 
\end{eqnarray}
}

For a bath with a spectrum that is stronger at low frequencies (\( \alpha \leq 2 \)),
the Green's function may decay to zero for \( t\rightarrow \infty  \), which
means that there is no delta peak any more in the density of states. 

The resulting single-particle density of states is presented in Fig.~\ref{Abb:dos}
for different values of the exponent \( \alpha  \) characterizing the strength
of the bath spectrum at low frequencies, both for zero and finite temperatures.

\begin{figure}
\centerline{\hbox{  \includegraphics[width=8cm]{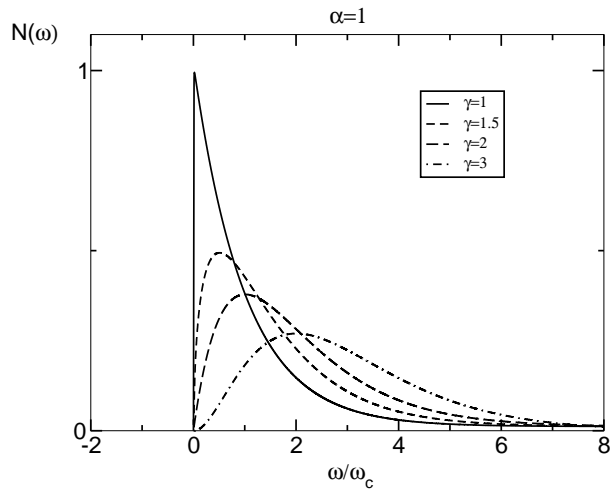}}}

\caption{\label{fig3}The density of states for Nyquist noise (\protect\( \alpha =1\protect \))
at \protect\( T=0\protect \), for different values of the power-law exponent
\protect\( \gamma \protect \). At finite temperature \protect\( T\protect \),
this goes over into a Lorentz peak of width \protect\( \propto T\protect \),
see text.}
\end{figure}
Here, we will analyze in more detail the case of Nyquist noise, with a linear
spectrum \( C(\omega )=C_{0}\omega  \) at small frequencies. 

If the temperature is finite, we may set \( n(\omega )\approx T/\omega  \)
for sufficiently low frequencies. The real part of the exponent (\ref{greenexponent})
then becomes, in the long-time limit:

\begin{equation}
-2\pi C_{0}\kappa ^{2}T\cdot t\, .
\end{equation}

The prefactor in this expression defines the decay rate for the Green's function,
which determines the width of the Lorentzian that arises in the density of states.

At zero temperature the thermal excitation of the bath oscillators vanishes,
so only the zero-point contribution remains. In that case, we have to specify
the behaviour of the bath spectrum at high frequencies, because it becomes important
even at \( t\rightarrow \infty  \): Choosing either a sharp cutoff \( C(\omega )=C_{0}\omega \theta (\omega _{c}-\omega ) \)
or an exponential decay \( C(\omega )=C_{0}\omega \exp \left( -\omega /\omega _{c}\right)  \)
results in the same long-time behaviour of the relevant integral: 

\begin{equation}
\int _{0}^{\infty }d\omega \, \frac{C(\omega )}{\omega ^{2}}\left( \cos \left( \omega t\right) -1\right) =-C_{0}\ln \left( \omega _{c}t\right) +h(t)\, .
\end{equation}

Here \( h(t) \) is a function that saturates to a constant value for \( t\rightarrow \infty  \).
The logarithm in the exponent leads to a power-law decay of the Green's function

\begin{equation}
\label{powerlawdecay}
G^{R}_{p}(t)\propto \left( \omega _{c}t\right) ^{-\gamma }\, ,
\end{equation}
 with an exponent \( \gamma  \) proportional to the strength of the bath, the
coupling and the momentum squared:

\begin{equation}
\gamma =\kappa ^{2}C_{0}=\left( \frac{gp}{m}\right) ^{2}C_{0}\, .
\end{equation}

The detailed behaviour of the lineshape (not only the linewidth) then depends
on \( \gamma  \), see Fig.~\ref{fig3}.

We will encounter such a power-law decay again in the discussion of dephasing
for the Nyquist bath. It is similar to the power-law behaviour found in the
model of Quantum Brownian motion \cite{2,3,schramm}. However, there are important
differences between the Nyquist bath considered here and the (stronger) Ohmic
bath. These will be discussed in section \ref{discussdephasing}.

\section{Many-particle problem}
\label{Sec:manyparticle}

Now we consider the situation of many electrons on the ring, in the grand canonical
ensemble at a given value of the chemical potential \( \mu  \) and at an arbitrary
temperature \( T=\beta ^{-1} \).

\subsection{Grand canonical partition sum and persistent current}

\label{partsumsec}As a preparation for the evaluation of various averages carried
out below, we need the grand canonical partition sum:

\begin{eqnarray}
Z_{gc}(\beta ,\mu ) & = & \sum _{N=0}^{\infty }e^{\beta \mu N}tr\left( e^{-\beta \hat{H}_{N}}\right) \nonumber \\
 & = & \sum _{N}e^{\beta \mu N}\sum _{\left\{ p_{j}\right\} }tr_{B}\left( e^{-\beta \hat{H}[\{p_{j}\}]}\right) \, .
\end{eqnarray}

Here we have used the notation \( \left\{ p_{j}\right\}  \) (which has been
introduced above) for a given configuration of \( N \) particles. \( tr_{B} \)
in the second line denotes the trace with respect to the bath oscillators. If
we introduce an additional static magnetic flux \( \Phi \), this corresponds
to a shift of the momenta \( p_{j}\mapsto p_{j}-e\Phi /Lc \) in all the
formulas given below, such that the partition sum becomes dependent on \( \Phi  \)
as well. 
Note that we assume the effects of the charging energy of the ring to be negligible here, unlike the treatment of cotunneling in subsection \ref{abcotunnel}.

The Hamiltonian \( \hat{H}[\{p_{j}\}] \) can be split into a part containing
only the momenta \( p_{j} \) and another one which represents bath oscillators
which are shifted (depending on \( p_{j} \)) but still have the original frequencies
(if we neglect the \( g^{2}\hat{\phi }^{2} \) term like above):

\begin{eqnarray}
\sum _{j=1}^{N}\frac{p_{j}^{2}-2gp\hat{\phi }}{2m}+\hat{H}_{bath} &  & \nonumber \\
=T[\{p_{j}\}]+\hat{H}_{bath}[\{p_{j}\}]\, , &  & 
\end{eqnarray}

where \( \hat{H}_{bath}[\{p_{j}\}] \) is derived from the original bath Hamiltonian
(\ref{hambath}) by a shift in oscillator coordinates

\begin{equation}
\hat{Q}_{l}\mapsto \hat{Q}_{l}-\frac{g}{mM\omega _{l}^{2}\sqrt{N_{osc}}}\sum _{j}p_{j}\, ,
\end{equation}

and \( T[\{p_{j}\}] \) is the residual {}``kinetic energy{}'' term that only
depends on the set of occupied momenta \( \left\{ p_{j}\right\}  \): 

\begin{equation}
T[\{p_{j}\}]\equiv \frac{1}{2m}\left( \sum _{j}p_{j}^{2}-\xi \left( \sum _{j}p_{j}\right) ^{2}\right) \, .
\end{equation}

Here we have used the constant factor \( \xi  \) defined in the discussion
of the effective mass, see Eq. (\ref{effectivemass}). \( T[\{p_{j}\}] \) differs
from the single particle case by the appearance of the total momentum \( P=\sum p_{j} \),
which makes it impossible to write this as a sum over the kinetic energies of
individual electrons with renormalized masses. Physically, all the electrons
are coupled to one and the same bath and this influences the center-of-mass
motion. The dependence of \( T[\{p_{j}\}] \) on the total momentum introduces
a kind of simple {}``effective interaction{}'' between the electrons, which
affects thermodynamic averages. For example, the average occupation number for
a given momentum does not follow the Fermi-Dirac distribution. Note that formally
the kinetic energy \( T[\{p_{j}\}] \) can become negative for large total momenta
\( P \) if \( \xi >1/N \). However, this only means that higher orders in
\( \xi  \) would have to be taken into account (neglecting \( g^{2}\hat{\phi }^{2} \)
becomes invalid), as discussed already in connection with the effective mass
for a single particle (see Eq. \ref{effectivemass}).

Using this definition, we can rewrite \( Z_{gc} \), taking into account that
the partition sum of the bath of shifted harmonic oscillators is equal to that
of the unperturbed bath, \( Z_{HO} \), and therefore does not depend on \( \left\{ p_{j}\right\}  \):

\begin{equation}
\label{zgcan}
Z_{gc}=Z_{HO}\sum _{N}e^{\beta \mu N}\sum _{\left\{ p_{j}\right\} }e^{-\beta T[\{p_{j}\}]}\, .
\end{equation}

The partition sum without the contribution due to the harmonic oscillators will
be denoted by \( \tilde{Z}_{gc} \) from now on:

\begin{equation}
\tilde{Z}_{gc}\equiv \frac{Z_{gc}}{Z_{HO}}\, .
\end{equation}
We can simplify (\ref{zgcan}) further by rewriting \( \tilde{Z}_{gc} \) in
the following way, using both the average \( \left\langle \cdot \right\rangle _{0} \)
and the partition sum \( \tilde{Z}_{gc}^{0} \) with respect to the system of
\emph{free} electrons:

\begin{eqnarray}
\tilde{Z}_{gc}=\sum _{N,\left\{ p_{j}\right\} }e^{-\beta \sum _{j}\left( p_{j}^{2}/2m-\mu \right) }e^{\beta \xi P^{2}/2m} &  & \nonumber \\
=\tilde{Z}_{gc}^{0}\, \left\langle e^{\beta \xi \hat{P}^{2}/2m}\right\rangle _{0}\approx \tilde{Z}_{gc}^{0}\, \left( 1+\beta \frac{\xi }{2m}\left\langle \hat{P}^{2}\right\rangle _{0}\right) \, . &  & \label{partitionsumZ} 
\end{eqnarray}

Here, we have kept only the lowest nonvanishing order in \( \xi  \), since
everything else would be inconsistent within the framework of our approximation. 

Now we can apply these results to the calculation of the persistent current
of our system of electrons coupled to the fluctuating flux.

The average current which flows for a given external magnetic flux \( \Phi  \)
is the derivative of the thermodynamic potential \( \Omega =-T\ln Z_{gc} \)
with respect to the flux itself:

\begin{eqnarray}
\left\langle \hat{I}\right\rangle &=&cT\frac{\partial }{\partial \Phi }\ln Z_{gc}  \nonumber \\
&=&cT\frac{\partial }{\partial \Phi }\left\{ \ln \tilde{Z}_{gc}^{0}+\beta \frac{\xi }{2m}\left\langle \left( \hat{P}-\hat{N}e\Phi /cL\right) ^{2}\right\rangle _{0}\right\}  \nonumber \\
&=&\frac{e}{L}\left\langle \hat{V}\left( 1-\xi \hat{N}\right) \right\rangle _{0}\, .  
\end{eqnarray}

Here \( \hat{V}=(\hat{P}-\hat{N}e\Phi /\left( cL\right) )/m \) is the
total velocity operator for the electrons. We have replaced \( \hat{P} \) in
(\ref{partitionsumZ}) by the expression valid in the presence of the external
static flux. Obviously, the persistent current is reduced by a factor of about
\( \left( 1-\xi N\right)  \). The result does not depend on the details of
the bath spectrum, only on \( \xi  \), which is well-behaved also for the case
of the Nyquist bath. Recall that for values \( \xi >1/N \) the approximation
used here is invalid, as explained above.

The magnitude of the reduction of the persistent current may be understood physically
in the following way: If one imagines suddenly switching on the external magnetic
flux, an electric field pulse will be produced, which, at first, freely accelerates
the electrons on the ring, leading to a current which is proportional to the
number \( N \) of electrons. This again produces a change in the magnetic flux
which prompts a reaction of the bath (e.g. the external coil producing the Nyquist
noise). The back-action onto the electrons deccelerates them, decreasing the
velocity of each electron by an amount proportional to \( N \) and depending
on the coupling strength between the ring and the external coil, which is contained
in \( \xi  \). This leads to the reduction factor \( 1-\xi N \) obtained above. 

Thus, in our model, the reduction of the persistent current is similar in its
origin to the appearance of an effective mass. This can be seen most clearly
by considering the special case of a {}``fast{}'' bath, whose spectrum has
a lower frequency cutoff or, at least, vanishes quickly with decreasing frequency.
It follows the motion of the electrons adiabatically and is not able to lead
to dephasing on long time scales (see the discussion of the two-particle Green's
function below, as well as the cotunneling setup discussed in the last section).
However, it still leads to a reduction of the persistent current, since the
quantity \( \xi  \) represents an integral over \emph{all} frequencies.

\subsection{Single-particle Green's function}

The single-particle Green's function is defined by

\begin{equation}
\label{greendef}
iG_{p}(t)=\left\langle \hat{\Psi }_{pt}\hat{\Psi }_{p0}^{\dagger }\right\rangle \theta (t)-\left\langle \hat{\Psi }_{p0}^{\dagger }\hat{\Psi }_{pt}\right\rangle \theta (-t)\, .
\end{equation}

Evaluation of both expectation values proceeds in the same way, so we will only
treat the first one here:

\begin{eqnarray}
\left\langle \hat{\Psi }_{pt}\hat{\Psi }_{p0}^{\dagger }\right\rangle = &  & \nonumber \\
Z_{gc}^{-1}\sum _{N}e^{\beta \mu N}tr\left( e^{-\beta \hat{H}_{N}}e^{i\hat{H}_{N}t}\hat{\Psi }_{p}e^{-i\hat{H}_{N}t}\hat{\Psi }_{p}^{\dagger }\right) \, . &  & \label{greeneval} 
\end{eqnarray}

The trace is evaluated by summing over all configurations \( \left\{ p_{j}\right\}  \)
of \( N \) particles and using the fact that the bath does not change a given
configuration. Therefore it is equal to:

\begin{equation}
\label{greenev2}
\sum _{\left\{ p_{j}\right\} }\left( 1-n_{p}[\{p_{j}\}]\right) tr_{B}\left( e^{-\beta \hat{H}[\left\{ p_{j}\right\} ]}e^{i\hat{H}[\{p_{j}\}]t}e^{-i\hat{H}[\{p_{j}\}']t}\right) \, .
\end{equation}

Here \( \left\{ p_{j}\right\} ' \) is the configuration with one particle added
in state \( p \), and the prefactor is zero whenever that state is already
occupied. Now we introduce the interaction picture with respect to \( \hat{H}[\{p_{j}\}] \).
Furthermore, we will use the partition sum for a given configuration,

\begin{equation}
Z_{\{p_{j}\}}=tr_{B}\left( e^{-\beta \hat{H}[\{p_{j}\}]}\right) =Z_{HO}\cdot e^{-\beta T[\{p_{j}\}]}\, ,
\end{equation}

in order to define the average over the bath oscillators which are shifted depending
on \( \left\{ p_{j}\right\}  \):

\begin{eqnarray}
\left\langle \ldots \right\rangle _{\left\{ p_{j}\right\} }\equiv Z_{\left\{ p_{j}\right\} }^{-1}tr_{B}\left( e^{-\beta \hat{H}[\left\{ p_{j}\right\} ]}\ldots \right)  &  & \nonumber \\
=Z_{HO}^{-1}tr_{B}\left( e^{-\beta \hat{H}_{bath}[\{p_{j}\}]}\ldots \right) \, . &  & 
\end{eqnarray}

(See the discussion of \( Z_{gc} \) in section \ref{partsumsec} for a definition
of the quantities \( Z_{HO} \) and \( \hat{H}_{bath}\left[ \left\{ p_{j}\right\} \right]  \))

With these definitions, expression (\ref{greenev2}) becomes:

\begin{eqnarray}
\label{exponavg}
\sum _{\left\{ p_{j}\right\} }Z_{\left\{ p_{j}\right\} }\left( 1-n_{p}[\{p_{j}\}]\right) & & \nonumber \\ 
\times \left\langle \hat{T}\exp \left[ -i\frac{p^{2}}{2m}t+i\frac{gp}{m}\int _{0}^{t}dt'\, \hat{\phi }(t')\right] \right\rangle _{\left\{ p_{j}\right\} }\, . & &
\end{eqnarray}

In evaluating the average of the exponential, we need the expectation value
of \( \hat{\phi } \) which does not vanish in this case, since the oscillators
are shifted by an amount proportional to the total momentum of the given configuration:

\begin{equation}
\label{phiavg}
g\bar{\phi }=g\left\langle \hat{\phi }\right\rangle _{\left\{ p_{j}\right\} }=\xi \left( \sum _{j}p_{j}\right) \, .
\end{equation}

Apart from this, we can proceed exactly as before in the strictly single-particle
case, see Eq. (\ref{timeorder}), in order to arrive at an exponent involving
the thermal time-ordered correlator of \( \hat{\phi } \) for the unperturbed
harmonic oscillators. Using this and (\ref{phiavg}) to evaluate (\ref{exponavg}),
we have arrived at the desired result for one half of the Green's function,
Eq. (\ref{greeneval}). Proceeding analogously for the other half and using
the definition (\ref{Kdef}) for \( K(\omega ;\kappa ) \), the Fourier transform
of the Green's function is given by:

\begin{eqnarray}
G_{p}(\omega )= &  & \nonumber \\
\tilde{Z}_{gc}^{-1}\sum _{\left\{ p_{j}\right\} }\exp \left( -\beta \left\{ \sum _{j}\left( \frac{p_{j}^{2}}{2m}-\mu \right) -\frac{\xi }{2m}P^{2}\right\} \right)  &  & \nonumber \\
\times \left\{ \left( 1-n_{p}[\{p_{j}\}]\right) K(\omega -\frac{p^{2}}{2m^{*}}+\frac{\xi }{m}pP;\frac{gp}{m})\right.  &  & \nonumber \\
\left. -n_{p}[\{p_{j}\}]\, K(-\omega +\frac{p^{2}}{2m}\left( 1+\xi \right) -\frac{\xi }{m}pP;\frac{gp}{m})\right\} \, . &  & \label{greenfctresult} 
\end{eqnarray}

The sum runs over all configurations (of any particle number \( N \)) and \( P=\sum p_{j} \)
is the total momentum. 

From this result one can see that the lineshape of the DOS in the many-particle
case is derived from the single-particle result \( K(\omega ;\kappa ) \). However,
at finite temperatures, it is the average over many such curves, each shifted
by an amount \( \propto pP \) that depends on the total momentum of the configuration.
Still, the linewidth (or lineshape) does not depend in any essential way on
the distance to the Fermi surface.

We will now discuss the temperature-dependence of the linewidth that results
from expression (\ref{greenfctresult}). The delta-peak which remains in \( K(\omega ) \)
at \( \omega =0 \) for a {}``weak bath{}'' (in this case: \( \alpha >2 \))
is smeared over a certain range due to the average over configurations with
different total momenta \( P \). In lowest order with respect to \( \xi  \),
we may neglect the dependence of the probability distribution on the coupling
to the bath. Then, the linewidth is obviously given by 

\begin{equation}
\delta \omega =\frac{\xi }{m}p\delta P\, ,
\end{equation}

where \( \delta P=\sqrt{\left\langle P^{2}\right\rangle _{0}} \) is the spread
in total momentum, calculated for the \emph{original} free-electron system.
We have \( \left\langle P^{2}\right\rangle _{0}=NmT \) and therefore a linewidth
which increases with the square root of \( T \):

\begin{equation}
\delta \omega \propto \xi \sqrt{T}\, .
\end{equation}

Note that the corresponding spread \( \delta p=\delta \omega /v \) in momentum
space is given by \( \xi \sqrt{NmT} \) and can very well exceed the distance
\( 2\pi /L \) of the quantized momenta, in spite of the restriction \( \xi N\ll 1 \)
(and also in spite of the restriction \( \sqrt{mT}\ll p_{F} \) for the degenerate
regime). Therefore, it is reasonable to speak of a linewidth, provided one does
not resolve the quantized level-structure on the ring.

\subsection{Dephasing: Two-particle Green's function}

While the decay of the single-particle Green's function in time is connected
with every interaction process that changes the state of the electron or brings
about random changes in its phase, it is not sufficient to know about this decay
if one asks about dephasing. After all, there are situations where an electron
interacts with a bath, such that its Green's function decays quickly but it is
still able to show an interference pattern. This will happen whenever the trace
left by the particle in the bath is not enough to decide which path it has gone,
so that the different possibilities still interfere \cite{sai}. Therefore,
one must ask about the time-evolution of the density matrix (and, in particular,
the decay of its off-diagonal elements) in order to study dephasing. Given
a small initial perturbation that creates a nonequilibrium situation, this time-evolution
is determined by the two-particle Green's function in a linear-response calculation.

The following calculation basically proceeds along the same lines as that given
in the previous section, so we will keep it brief. For our purposes, we do not
need the two-particle Green's function for arbitrary values of the four time-arguments,
but only for a perturbation acting at time \( 0 \) and a density matrix evaluated
at time \( t \). Furthermore, since the bath does not change the occupation
of momentum states, the only nontrivial contribution arises from the following
product of four electron operators, in which only two momenta \( p \) and \( p' \)
appear. \( p \) refers to the hole that is created by the perturbation while
\( p' \) belongs to the electron.

\begin{eqnarray}
\left\langle \hat{\Psi }_{pt}^{\dagger }\hat{\Psi }_{p't}\hat{\Psi }_{p'0}^{\dagger }\hat{\Psi }_{p0}\right\rangle  &  & \nonumber \\
=Z_{gc}^{-1}\sum _{N}e^{\beta \mu N}tr\left( e^{-\beta \hat{H}_{N}}e^{i\hat{H}_{N}t}\hat{\Psi }_{p}^{\dagger }\hat{\Psi }_{p'}e^{-i\hat{H}_{N}t}\hat{\Psi }_{p'}^{\dagger }\hat{\Psi }_{p}\right) \, . &  & \label{twopartgr1} 
\end{eqnarray}

Inserting an appropriate basis of system states \( \left| \left\{ p_{j}\right\} \right\rangle  \)
, using the interaction picture with respect to \( \hat{H}[\{p_{j}\}] \) and
carrying out the average of the exponential in the usual way, we arrive at the
following result for the (half-sided) Fourier transform of Eq. (\ref{twopartgr1}):

{\footnotesize \begin{eqnarray}
\tilde{Z}_{gc}^{-1}\sum _{\left\{ p_{j}\right\} }\exp \left( -\beta \left\{ \sum _{j}\left( \frac{p_{j}^{2}}{2m}-\mu \right) -\frac{\xi }{2m}P^{2}\right\} \right)  &  & \nonumber \\
\times n_{p}[\{p_{j}\}](1-n_{p'}[\{p_{j}\}])\,  &  & \nonumber \\
\times iK(\omega -\frac{p'^{2}-p^{2}}{2m}+\frac{\xi }{2m}\left( p'-p\right) ^{2}+\frac{\xi }{m}(p'-p)P;\frac{g(p'-p)}{m})\, . &  & \label{twopartgr2} 
\end{eqnarray}}

The notation is the same as for Eq. (\ref{greenfctresult}). The most important
difference consists in replacing \( p \) by \( p'-p \) in the factor \( \kappa =gp/m \)
which determines the strength of the decay.

For the Nyquist case, we thus obtain a finite dephasing rate

\begin{equation}
\label{tauphiexactnyquist}
\frac{1}{\tau _{\varphi }(p,p')}=2\pi TC_{0}\left( \frac{g}{m}\right) ^{2}\left( p'-p\right) ^{2}\, ,
\end{equation}
 which is proportional to the difference in momenta squared, the bath and coupling
strengths and the temperature. The change in phase brought about by the fluctuating
flux is proportional to \( p \), so the phase-difference, whose variance appears
in the exponent, goes like \( p'-p \). To avoid confusion, we emphasize that
there is no universally applicable definition of a {}``dephasing rate{}''.
In our case, we use this term to refer to the exponential decay of the two-particle
Green's function as introduced above. Note that the dephasing rate can be small
even if the decay rates associated with the (single-particle) Green's functions
of the individual states are large. This is an example of the general behaviour
mentioned above: It is reasonable that dephasing is strongest whenever the momenta
of the two states, whose superposition is examined, differ widely. Then, the
bath, which couples to the momentum, can easily distinguish between these states
even after a short time. 

The dephasing rate \( \tau _{\varphi }^{-1} \) vanishes at \( T=0 \). However,
even at \( T=0 \) the off-diagonal element of a density matrix which initially
describes a coherent superposition between momentum states \( p \) and \( p' \)
decays to zero completely in the limit \( t\rightarrow \infty  \). This decay
proceeds with a power-law, as we have already observed for the single-particle
Green's function, see Eq. (\ref{powerlawdecay}). In the situation considered
here, the exponent \( \gamma  \) is equal to

\begin{equation}
\label{gamma2part}
\gamma =\left( \frac{g}{m}\right) ^{2}\left( p'-p\right) ^{2}C_{0}\, .
\end{equation}

If the bath is sufficiently weak at low frequencies (\( C(\omega )=C_{0}\omega ^{\alpha } \)
with \( \alpha >2 \)), the decay of the off-diagonal elements in time saturates
at a finite value, in contrast to the Nyquist case discussed above. Then the
dephasing rate, defined as the prefactor of \( t \) in the exponential decay-law,
is strictly zero, even at finite temperatures. This behaviour is related to
the diagonal coupling between system and bath. If a nondiagonal coupling
were introduced, there would be transitions from excited electronic states towards
lower ones, accompanied by the spontaneous emission of a bath phonon. In that
case, the decay rate in the single-particle Green's function and the dephasing
rate would be nonzero also at \( T=0 \) but strongly dependent on the distance
to the Fermi surface, due to the suppression of the density of final states
for such transitions brought about by the Pauli-principle. Therefore, in a simple
Golden rule calculation, these decay rates would vanish at \( T=0 \) when one
approaches the Fermi surface.

For \( \alpha =2 \), we find power-law dephasing only at finite temperatures,
with an exponent proportional to \( T \).

\subsection{\label{discussdephasing}Discussion of dephasing for the {}``Nyquist bath{}''}

The Nyquist bath is characterized by a fluctuation spectrum of flux and vector
potential which is linear in \( \omega  \) (at zero temperature), therefore
leading to a spectrum for the electric field that behaves like \( \omega ^{3} \).
This is exactly the spectrum of the zero-point fluctuations of the electric
field in the vacuum. The main distinction between those fluctuations and the
Nyquist noise considered here is that the latter leads to a force which is homogeneous
around the ring and therefore is compatible with the translational invariance
of our one-dimensional system of electrons. Furthermore, its magnitude depends
on the geometry and resistance of the external circuit producing the equilibrium
current noise. Apart from these differences, we can use our understanding of
the electromagnetic vacuum fluctuations to discuss the effects of the Nyquist
bath in a qualitative manner. 

In particular, free ballistic motion is not affected, since the radiation reaction
force only acts on an accelerated charge. Therefore, the populations of the
electronic momentum eigenstates do not decay, as we have already observed. This
is in contrast to the effect of a type of bath that leads to velocity-proportional
friction, for example. The system is not ergodic, since the memory of the initial
conditions is not lost completely. In a basis other than the momentum basis,
the off-diagonal elements of the density matrix show only partial decay. That
there \emph{is} some decay of the coherences is connected to the smearing of
the position of the particle in the course of time. Usually, this effect is
neglected in the discussion of dissipative quantum motion of a free particle
under the influence of a bath corresponding to electromagnetic vacuum fluctuations, since the ballistic
expansion of an initially localized ensemble of particles dominates \cite{schramm}. In our
case, it is important, since, for example, a superposition of two counterpropagating
plane waves on the ring will first form a standing wave pattern, whose visibility
then gradually decreases. The fact that, at \( T=0 \), the decay of the visibility
proceeds as a power-law can be understood most easily from the results of an
old semiclassical analysis of the Lamb-shift due to Welton \cite{welton,milonni}: The vacuum fluctuations of the electric field lead to a jitter
of the electron position, such that the variance \( \left\langle \delta x^{2}\right\rangle  \)
of its coordinate is given by the logarithm containing the ratio of an upper
cutoff-frequency (there taken to be the Compton frequency) and a lower cutoff
(the characteristic frequency of electron motion around the nucleus). In our
case, the lower cutoff frequency actually is given by the inverse of the observation
time, such that \( \left\langle \delta x^{2}\right\rangle \propto \ln \left( \omega _{c}t\right)  \).
For a superposition of plane waves of momenta \( \pm k \) on the ring, the
density matrix in position space contains cross-terms like \( \exp \left( i2kx\right)  \),
which, if averaged over \( \delta x \), give rise to a suppression factor \( \exp \left( -2k^{2}\left\langle \delta x^{2}\right\rangle \right) \propto t^{-\gamma } \).
This leads to complete decay of the interference pattern even at \( T=0 \).
Note, however, that here we have been considering a superposition of excited
states of the system and the decay of its coherences. In other problems of dephasing,
such as those encountered in weak localization, one usually discusses the limit
of zero-frequency response of the system to a small perturbation. A situation
which comes closer to this kind of question will be discussed in the next section.

The goal of the present work has been to analyze thoroughly a model situation
which shows some of the features important for dephasing at low temperatures,
not to propose some experimental measurement setup. Still, we will now briefly
discuss the expected magnitude of the effect due to Nyquist noise in an external
current coil. If the equilibrium current fluctuations are produced by an external
coil whose circumference is similar to that of the Aharonov-Bohm ring ($L$) and which
is placed about a distance $L$ away, the dephasing rate is estimated to be

\begin{equation}
\frac{\hbar}{\tau _{\varphi }}\sim \left( \frac{e^{2}}{\hbar c}\right) ^{2}\left( \frac{v_{F}}{c}\right) ^{2}\frac{k_{B}T}{R/R_{K}}\, .
\end{equation}

Here \( v_{F} \) is the Fermi velocity on the ring, \( R \) is the resistance
of the external coil and \( R_{K}=h/e^{2} \) is the quantum of resistance.
Both the square of the fine-structure constant in front of the expression and
the ratio of the Fermi velocity to the speed of light render the effect very
small under reasonable experimental conditions.

Note that the fluctuations of the vacuum \emph{magnetic} field have a weaker
power-spectrum, which leads to \( \left\langle AA\right\rangle _{\omega }\propto \omega ^{3} \)
at \( T=0 \), instead of \( \left\langle AA\right\rangle _{\omega }\propto \omega  \).
The vacuum fluctuations of the \emph{electric} field, however, do lead to a
linear spectrum in the vector potential fluctuations (see discussion above).
On the other hand, the electric field (at large wavelengths) is homogeneous
only in free space, not with respect to its projection onto the ring, where
it has a position dependence \( \cos \left( 2\pi x/L\right)  \). Thus, the
coupling is not diagonal in the momentum basis and is not included in our model.
If one estimates the order of magnitude of the corresponding {}dephasing
rate{} (inelastic transition rate), one arrives at \( \tau _{\varphi (em)}^{-1}\sim \left( e^{2}/\hbar c\right) \left( v_{F}/c\right) ^{2}\left( v_{F}/L\right)  \).
Similarly, one may estimate the strength of fluctuations due to shot-noise of
the external current. In a situation where the external coil producing a static
magnetic flux on the order of \( \sim \Phi _{0} \) is identical with that where
the Nyquist noise originates, this leads to an effective dephasing rate of
\( \tau _{\varphi (shot)}^{-1}\sim \left( e^{2}/\hbar c\right) \left( v_{F}/c\right) v_{F}/L \),
which may be much larger than that due to the Nyquist noise. Note that shot
noise cannot be described by our model, since it is a nonequilibrium phenomenon and cannot be represented
by the usual bath of harmonic oscillators at low temperatures. However, the effects
of shot noise would be reduced in a different geometry where a larger current
(with correspondingly smaller \emph{relative} magnitude of the shot noise) produces
the same static magnetic flux through the Aharonov-Bohm ring.

\subsection{Relevance of the term quadratic in the flux}

\label{quadraticfluxterm}In all of the preceding calculations, we have neglected
the term \( g^{2}\hat{\phi }^{2} \) which appears in the kinetic energy of
the particle but does not couple to the momentum. This approximation has been
necessary to use the well-known formula \( \left\langle \exp \left( iX\right) \right\rangle =\exp \left( -\left\langle X^{2}\right\rangle /2\right)  \)
for a Gaussian random variable \( X \) in the classical case or the analogous
expression derived from Wick's theorem in the quantum case. \( \hat{\phi }^{2} \)
is quadratic in the coordinates of the bath oscillators, whereas \( \hat{\phi } \)
is a linear bosonic variable (Gaussian random variable in the classical case).
If the term \( g^{2}\hat{\phi }^{2}/2m \) is kept in the Hamiltonian, the eigenfrequencies
and normal coordinates of the bath of harmonic oscillators are changed by an
amount that depends on \( g \) but not on the particle momentum \( p \). This
becomes important at larger values of \( g \), where the effective mass turns
out to be \( m^{*}=m(1+\xi ) \) instead of the value \( m^{*}=m(1-\xi )^{-1} \)
derived without the \( \hat{\phi }^{2} \)-term, see Eq. (\ref{effectivemass}).
Note that this difference persists also in the thermodynamic limit \( N_{osc}\rightarrow \infty  \).
However, the qualitative behaviour of the bath spectrum at low frequencies is
not changed for the spectra \( C(\omega )\propto \omega ^{\alpha } \) with
\( \alpha \geq 1 \) that have been considered here. Therefore, this term is
unimportant for the qualitative conclusions about dephasing, although it can
change quantitative results for larger \( g \) and does change the single-particle
Green's function. The latter involves a change in particle number, so that an
additional \( \hat{\phi }^{2} \) is introduced into the Hamiltonian.

\section{Aharonov-Bohm interference in cotunneling through the ring}
\label{abcotunnel}
\begin{figure}
\centerline{\hbox{ \includegraphics[width=6cm]{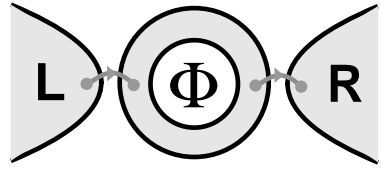} }}

\caption{The tunneling setup discussed in the text. }

\label{tunnelsetup}
\end{figure}

In the following, we will discuss the influence of the bath on the Aharonov-Bohm
(AB) effect, i.e. on the flux-dependence of the transport current through the
ring. We consider tunneling into and out of the ring, taking place at two electrodes
to the left and right of the ring (see Fig.~\ref{tunnelsetup}). A tunneling
situation is the appropriate one for our model, since attaching current leads
would severely alter the system. Note that independent (sequential) tunneling
only probes the density of states at the two contacts. Therefore, in order to
observe the AB interference effect, we have to consider a Coulomb blockade situation,
in which any electron tunneling into (or out of) the ring will enhance the total
energy by the charging energy of the ring which is much larger than
the bias voltage \( V \) and the temperature \( T \). In such a case, transport
through the ring is possible only via cotunneling\cite{averinnazarov}, i.e. a two-step process involving
a virtual intermediate state belonging to a different number of electrons on
the ring. A strong dependence of the tunneling current on the external magnetic
flux, with a complete suppression at \( \Phi _{0}/2 \) due to destructive interference,
is visible only in the {}``elastic cotunneling{}'' contribution, where the
electronic state of the ring is left unaltered in the process. It is linear
in the bias voltage and will dominate the inelastic contribution at low temperatures
and for small bias voltages (see the discussion at the end of this section and
Ref.~\onlinecite{averinnazarov}).

Now consider cotunneling at \( T=0 \) under the influence of the bath. The
semiclassical analysis of Aharonov-Bohm interference given in a preceding section
is not applicable for \( V\rightarrow 0 \), since it assumes the electron can
emit or absorb an arbitrary amount of energy. In the quantum-mechanical
calculation, suppression of interference is due to the electron leaving a trace
in the bath that permits, at least in principle, to decide which of the two
arms of the ring the electron has traveled. This involves a transfer of energy
between electron and bath. The bath spectrum determines the amount of bath oscillators
able to absorb the small energy \( \leq eV \) which can be emitted by the electron.
Therefore, it is to be expected that dephasing at zero temperature is suppressed
for \( V\rightarrow 0 \) due to the energy conservation constraint. This will
be confirmed by the calculation described in the following, although there are
renormalization effects that change the strength of the tunneling current away
from the point of perfect destructive interference, \( \Phi =\Phi _{0}/2 \). 

\begin{figure}
\centerline{\hbox{ \includegraphics[width=8cm]{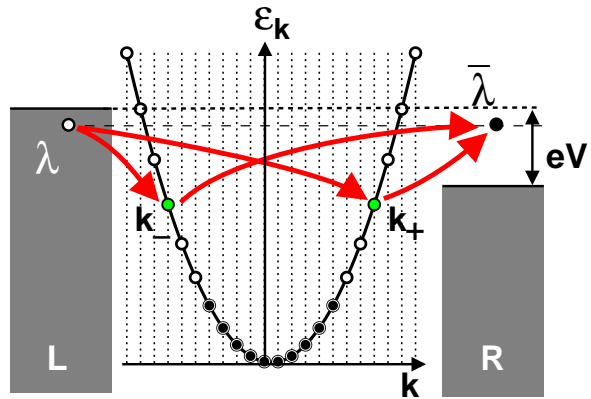} }}

\caption{\label{energydiag}Energy diagram for cotunneling through the AB-ring, at \protect\( \Phi =\Phi _{0}/2\protect \).
The initial, final and two possible intermediate states are indicated (see main
text). The charging energy has to be added to the single-particle energy of
the intermediate state shown here.}
\end{figure}

The tunneling process starts from a situation in which the ring is occupied
by the equilibrium number of electrons (depending on the value of a gate voltage)
and the Fermi seas in the left and right electrode are filled up to Fermi energies
that differ by the bias voltage, \( eV \), see Fig.~\ref{energydiag}. Throughout the following discussion,
we will assume the Fermi energy in the left electrode to be the larger one of
the two (and \( eV>0 \)). In the final state, an electron has appeared above
the right Fermi sea, leaving behind a hole in the left electrode. Although we
want to consider the situation where the electronic state of the ring has not
changed in the end, the final state of the bath may be different. The intermediate
state is characterized by an extra electron (or extra hole) present on the ring
and some arbitrary state of the bath. Using standard second-order Fermi's Golden
rule, the tunneling rate is obtained by summing over all intermediate states
(dividing by the proper energy denominator) and all final states whose energy
equals the initial energy:

\begin{equation}
\label{goldenrule}
\Gamma =2\pi \sum _{f}\left| \sum _{\nu }\frac{H^{T}_{f\nu }H^{T}_{\nu i}}{E_{\nu }-E_{i}}\right| ^{2}\delta \left( E_{f}-E_{i}\right) \, .
\end{equation}

Here \( \hat{H}^{T}=\hat{T}^{L}+\hat{T}^{R} \) is the sum of the tunneling
Hamiltonians belonging to the left and the right junction, while the energies
and eigenstates refer to the unperturbed Hamiltonian that includes everything
besides tunneling. In particular, it includes the coupling between electrons
and the bath, as well as the kinetic energies of electrons in the electrodes. 

At this point we would like to emphasize that using Fermi's Golden rule 
for the calculation of the cotunneling current does not in itself mean
taking into account the interaction between bath and system only in a
perturbative way. The intermediate states
being summed over in Eq.~(\ref{goldenrule}) are the exact eigenstates  of 
the full system of electrons on the ring coupled to the fluctuating flux.
In this sense, the coherence properties of the ring 
as a whole (including the bath) are 
tested by the cotunneling process. Applying the Golden rule 
in this context is roughly comparable to
using the Kubo formula in a linear-response calculation of, e.~g., 
the weak-localization magnetoconductance, which does not automatically imply a perturbative
description of the dephasing processes either.
Still, there
is an important difference: In our context, we essentially deal with a
scattering situation, such that energy conservation holds at least for
the initial and final states of the complete process. This will be seen
to be important for our conclusions about the strength (or absence of)
dephasing. What is neglected in our calculation are any equilibrium correlations 
between the state of the electrons in the leads and the state of the ring as a whole (including the fluctuating flux).


Before performing the calculation in the presence of the bath, we will briefly
describe how the destructive interference at \( \Phi _{0}/2 \) appears in this
formula, in the situation without fluctuating flux. In such a case, the intermediate
state \( \nu  \) refers solely to the electronic state \( k \) on the ring,
which is occupied by the additional electron in the course of tunneling. The
final state \( f \) is determined both by the state \( \lambda  \), which
is unoccupied in the left electrode after the tunneling process, and the state
\( \bar{\lambda } \), where the electron ends up in the right electrode (see
Fig.~\ref{energydiag}).

For simplicity, we will assume tunneling to take place only between two points,
for example from a point \( y_{L} \) at the tip of the left electrode to an
adjacent point \( x_{L} \) on the ring,

\begin{equation}
\label{eq50}
\hat{T}^{L}=t_{L}\hat{\Psi }^{\dagger }\left( x_{L}\right) \hat{\Psi }\left( y_{L}\right) +h.c.\, ,
\end{equation}

and likewise for the right electrode. \( t_{L} \) is a complex-valued tunneling
amplitude. Such a description will be appropriate as long as the extent of the
relevant region in which tunneling can take place is less than a wavelength.

The sum over intermediate electronic states \( k \) on the ring then contains
the following contribution which describes an electron going onto the ring from
the left electrode and leaving through the right electrode:

\begin{equation}
\label{tltr}
t_{L}t_{R}^{*}\Psi _{\lambda }(y_{L})\Psi _{\bar{\lambda }}^{*}(y_{R})\sum _{k}\frac{\Psi _{k}(x_{R})\Psi _{k}^{*}(x_{L})}{\epsilon _{k}+E_{C}-\epsilon _{\lambda }}\, .
\end{equation}

Here the sum over \( k \) is to be taken only over unoccupied single-electron
states on the ring. \( \Psi  \) refers to single-electron wavefunctions on
the ring and on the electrodes. Apart from the contribution listed here, there
is another, completely analogous, contribution which belongs to the situation
with an extra \emph{hole} on the ring in the intermediate state\cite{averinnazarov}.
Note that for the purposes of our discussion we will not distinguish between the
charging energies belonging to the electron- and hole-processes (assuming them to
be of about the same magnitude).

Perfect destructive AB-interference at an external static magnetic flux of \( \Phi =\Phi _{0}/2 \)
arises only for an even number of electrons on the ring. In this case, the energies
\begin{equation}
\epsilon _{k}=\frac{1}{2m}\left( k-\frac{2\pi }{L}\frac{\Phi }{\Phi _{0}}\right) ^{2}
\end{equation}

of the unoccupied states are pairwise degenerate, for \( k_{+}\equiv n2\pi /L \)
and \( k_{-}\equiv \left( 1-n\right) 2\pi /L \), see Fig.~\ref{energydiag}.
Therefore, the energy denominators for \( k_{+} \) and \( k_{-} \) are the
same, while the wavefunctions in the numerators produce a phase shift of \( \exp \left( i\left( k_{+}-k_{-}\right) L/2\right) =-1 \)
between the two possibilities, leading to complete cancellation of all terms
in the sum. The same applies to the sum over occupied states (for the situation
with an extra hole in the intermediate state).

Dephasing will, in general, {}``wash out{}'' this perfect destructive interference.
After taking the modulus squared of the sum of amplitudes given above in Eq.
(\ref{tltr}), which we briefly denote by \( A(k) \), one obtains {}``classical{}''
probabilities like \( \left| A(k_{+})\right| ^{2} \) but also cross-terms of
the form \( A^{*}(k_{+})A(k_{-}) \). A bath coupling to the electronic motion
will affect these terms differently, if it is able to {}``distinguish{}''
between the momenta \( k_{+} \) and \( k_{-} \). Usually, the cross-terms
are suppressed. Then, the different contributions cannot cancel any more. Note
that, away from perfect destructive interference, we have to expect an influence
of the bath on the magnitude of the tunneling current under any circumstances,
since mere renormalization effects like a change in the effective mass of the
electrons will be important. This is why we concentrate on the special case
of \( \Phi _{0}/2 \). Even in that situation, the interference minimum could
vanish in a rather trivial way due to renormalization effects, if one chose
a bath coupling asymmetrically to the two arms of the ring (or rather to left-
and right-going momenta), thereby leading to different transmission probabilities.
This is not the case in our model. 

\begin{figure}[htb]
\centerline{\hbox{  \includegraphics[width=5cm]{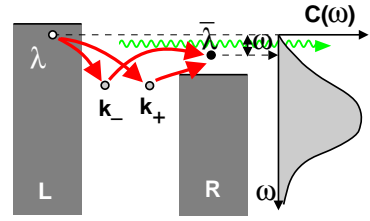} }}\nopagebreak
\caption{\label{phononemission}Cotunneling in the presence of the fluctuating flux:
Emission of a bath {}``phonon{}'' of frequency \protect\( \omega \protect \)
leads to an incoherent contribution to the tunneling current, where the destructive
interference between the two paths shown here is lost. The probability of emission
depends on the bath spectrum \protect\( C(\omega )\protect \). }
\end{figure}

At this point, we can give a simple counting argument to arrive quickly at the
voltage-dependence of the {}``incoherent{}'' contribution to the cotunneling
rate \( \Gamma  \), which is produced by the fluctuating flux and leads to
a nonvanishing current at \( \Phi _{0}/2 \). The sum over initial electronic
states on the left electrode is carried out over a region of extent \( eV \).
The probability of emission of a bath phonon is proportional to the bath spectrum
\( \propto \omega ^{\alpha } \), and we have to integrate this from \( 0 \)
to the maximum energy of the electron, which is again of order \( eV \). This
yields a voltage dependence \( \Gamma \propto V^{\alpha +2} \) of the incoherent
contribution to the tunneling current (see Figs. \ref{phononemission} and \ref{cotunnelrate}).
However, we want to analyze this in the following using a different, nonperturbative
scheme, thereby making contact to the Feynman-Vernon influence functional formalism\cite{1},
which is the {}``workhorse{}'' of many dephasing calculations.

To evaluate Eq.~(\ref{goldenrule}) in the presence of the bath, we rewrite
the sum over intermediate states as a time-integral:

\begin{eqnarray}
 &  & \sum _{\nu }\frac{H^{T}_{f\nu }H^{T}_{\nu i}}{E_{\nu }-E_{i}}=i\int _{0}^{\infty }dt\langle f|\hat{H}^{T}e^{-i(\hat{H}-E_{i})t}\hat{H}^{T}|i\rangle \label{evaltunneling}\\
 & = & i\int _{0}^{\infty }dt\sum _{k}T^{R}_{\bar{\lambda }k}T_{k\lambda }^{L}\langle f^{B}|e^{-i(\hat{H}[k]-E_{0})t}|0^{B}\rangle e^{-i(E_{C}-\epsilon _{\lambda })t} \, . \nonumber 
\end{eqnarray}

In the second line, which replaces Eq.~(\ref{tltr}) in the presence of the
bath, we have split off the contribution due to the electronic states. This
has been possible because only the tunneling operators can change the electronic
state, while the bath couples \textit{diagonally} to the electronic momentum
eigenstates. Furthermore, we have confined ourselves to the process with an
extra \textit{electron} in the intermediate state, a completely analogous contribution
for an extra \textit{hole} has to be added. \( \hat{H}[k] \) is the Hamiltonian
for a given configuration consisting of an extra occupied state \( k \) over
the original Fermi sea on the ring. It only acts on the bath Hilbert space,
where \( 0^{B} \) refers to the ground state of the bath prior to the tunneling
event and \( f^{B} \) is an arbitrary final state which the bath goes into
after the cotunneling process is finished. In our notation, the sum of electronic
kinetic energies is included in \( \hat{H}[k] \) as well, whereas the charging
energy \( E_{C} \) has been taken into account separately. The matrix elements
of the tunneling Hamiltonians \( T^{L,R} \) are taken between the electronic
states \( \bar{\lambda } \), \( k \) and \( \lambda  \) (compare Eqs. (\ref{eq50})
and (\ref{tltr})). \( E_{0} \) is the ground-state energy of the ring, including
the bath.

After taking the modulus squared of the sum given above, we arrive at the following
contribution to the tunneling rate \( \Gamma  \) at zero temperature:

\begin{eqnarray}
 &  & 2\pi \sum _{\lambda ,\bar{\lambda },k^{>},k^{<}}(T^{R}_{\bar{\lambda }k^{>}}T^{L}_{k^{>}\lambda })(T^{R}_{\bar{\lambda }k^{<}}T^{L}_{k^{<}\lambda })^{*} \\
 & \times  & \int ^{\infty }_{0}d\tau ^{>}e^{-i(E_{C}-\epsilon _{\lambda }-E_{0})\tau ^{>}}\int ^{\infty }_{0}d\tau ^{<}e^{+i(E_{C}-\epsilon _{\lambda }-E_{0})\tau ^{<}}\nonumber \\
 & \times  & \sum _{f^{B}}\langle \chi ^{<}(\tau ^{<})|f^{B}\rangle \delta (E_{f^{B}}-E_{0}-(\epsilon _{\lambda }-\epsilon _{\bar{\lambda }}))\langle f^{B}|\chi ^{>}(\tau ^{>})\rangle \; \nonumber, 
\end{eqnarray}

where \( k^{>(<)} \) denote unoccupied states on the ring. There are three
analogous contributions besides the one shown here, in which the tunneling takes
place in a different order (e.g. the process may start by an electron tunneling
out of the ring, leaving a hole behind, etc.). Note that a similar
expression arises in the derivation of the ``$P(E)$''-theory of a tunnel junction coupled to a 
dissipative bath\cite{ingoldnazarov,gschoen}.

In the preceding formula, the last line can be viewed as a kind of {}``\emph{generalized
influence functional}{}'' \( F[\tau ^{>},\tau ^{<},\omega =\epsilon _{\lambda }-\epsilon _{\bar{\lambda }}] \).
It is equal to the overlap between bath states \( \chi ^{>(<)} \) which have
been time-evolved out of \( 0^{B} \) under the action of \( \hat{H}[k^{>(<)}] \)
for some time \( \tau ^{>(<)} \). In contrast to the usual influence functional,
the time of evolution may be different for the two states and the overlap is
taken only with respect to bath states at an excitation energy \( \omega  \)
(which must equal the energy emitted by the electron). This difference is due
to the fact, that in our problem the energy conservation constraint must be
taken care of, since the electron cannot transfer an arbitrary amount of energy
to the bath. This clearly shows why a single-particle calculation using the usual
influence functional must fail when the amount of energy available is limited
due to low temperatures or low bias voltages. This problem has also been discussed
in Ref.~\onlinecite{cohenimry}, where the authors have used physical arguments to drop
certain {}``zero-point{}'' contributions to the dephasing rate obtained in
a single-particle calculation. The normal influence functional is recovered
by integrating over all possible energy transfers and setting \( \tau ^{>}=\tau ^{<} \):

\begin{equation}
F\left[ \tau \right] =\left\langle \chi ^{<}\left( \tau \right) \right. \left| \chi ^{>}\left( \tau \right) \right\rangle =\int _{-\infty }^{+\infty }d\omega F\left[ \tau ,\tau ,\omega \right] \, .
\end{equation}

If no bath is present or its spectrum has a lower cutoff which is larger than
the energy available to the electron, energy conservation leads to \( \omega \equiv 0 \),
such that the final and initial bath states coincide: \( f^{B}\equiv 0^{B} \).
Then, \( F \) is a product of a factor depending only on \( k^{>} \) and another
one, depending only on \( k^{<} \). In this case, the sums over \( k^{>(<)} \)
may be carried out separately, like before, and the terms will cancel again
(for \( \Phi _{0}/2 \)), provided the bath couples equally to \( k_{+} \)
and \( k_{-} \) (see discussion above). Although there is definitely no dephasing
in this case, the magnitude of the tunneling current may be changed for \( \Phi \ne \Phi _{0}/2 \),
due to the afore-mentioned renormalization effects.

The Fourier transform (in \( \omega  \)) of the generalized influence functional
may be written as follows: \begin{equation}
\label{geninfl}
F[\tau ^{>},\tau ^{<},\tau ]={e^{+iE_{0}\tau }\over 2\pi }\langle \chi ^{<}(\tau ^{<})|e^{-i\hat{H}[\emptyset]\tau }|\chi ^{>}(\tau ^{>})\rangle \; .
\end{equation}

\begin{figure}[htb]
\nopagebreak
\centerline{\hbox{ \includegraphics[width=5cm]{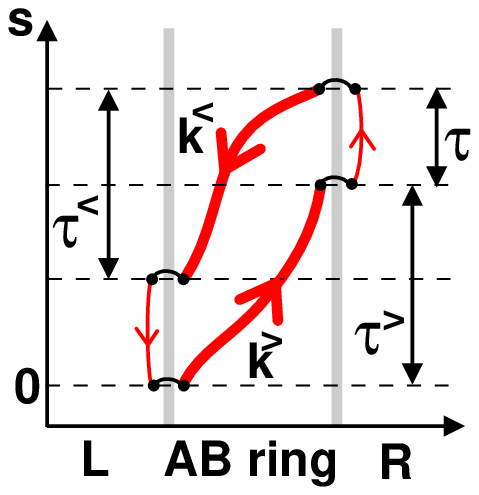} }} 
\caption{\label{keldyshcontour}Schematic {}``space-time{}'' diagram showing the Keldysh
contour which runs from \protect\( 0\protect \) to \protect\( \tau \protect \)
and back again. The interaction operator which couples to the bath is nonvanishing
only when there is an extra electron on the ring (thick lines), either in state
\protect\( k^{>}\protect \) or \protect\( k^{<}\protect \) (see main text).
Tunneling processes are indicated. }
\end{figure}

It can be represented as a Keldysh time-ordered expectation value \cite{keldysh,zagoskin},
apart from a prefactor \( \exp (-iE_{0}(\tau ^{>}-\tau ^{<})) \):

\begin{equation}
\label{overlap}
\langle \hat{T}_{K}e^{-i\oint _{K}ds\hat{V}_{I}(s)}\rangle _{0^{B}}\; .
\end{equation}
 Here, \( \hat{V}_{I}(s)\equiv g{\tilde{k}}\hat{\phi }(s)/m+{\tilde{k}}^{2}/(2m) \),
with \( {\tilde{k}}\equiv k-2\pi \Phi /({\Phi _{0}L}) \), where \( \Phi  \)
is the additional \emph{static} flux. \( \hat{V}_{I} \) couples the additional
electron in state \( k \) to the bath (and also incorporates the kinetic energy).
We have \( k=k^{>} \) if \( s \) is on the forward time-branch and \( 0\leq s\leq \tau ^{>} \),
while \( k=k^{<} \), if \( s \) is on the backward time-branch and \( \tau +\tau ^{>}-\tau ^{<}\leq s\leq \tau +\tau ^{>} \).
For all other times, \( \hat{V}_{I} \) vanishes. This is represented graphically
in Fig.~\ref{keldyshcontour}. Note that \( \hat{V}_{I} \) is taken in the
interaction picture with respect to the bath coupled to the original Fermi sea
on the ring. Once more, we have neglected the term \( g^{2}{\hat{\Phi }}^{2} \)
which turns out to be unimportant for the bath spectra considered below. 

Using Keldysh time-ordering and a linked cluster expansion (Wick's theorem),
we can represent Eq.~(\ref{overlap}) as an exponential containing double time-integrals
involving the Keldysh-time ordered correlation function of the bath operator
\( \hat{\phi } \): 

\begin{eqnarray}
\exp \left[-i\oint _{K}ds\left\langle \hat{V}_{I}(s)\right\rangle \right] & & \\
\times \exp \left[ -\frac{1}{2}\oint _{K}ds_{1}\oint _{K}ds_{2}\left\langle \hat{T}_{K}\delta \hat{V}_{I}(s_{1})\delta \hat{V}_{I}(s_{2})\right\rangle \right] & & \, \nonumber.
\end{eqnarray}

We have set \( \delta \hat{V}_{I}\equiv \hat{V}_{I}-\left\langle \hat{V}_{I}\right\rangle =g\tilde{k}\hat{\phi }/m \).
The principal steps involved in the evaluation of this expression are demonstrated
in Appendix \ref{Sec:influencederivation}, where we show how the usual Caldeira-Leggett
influence functional for a bath of harmonic oscillators can be derived very
efficiently using this method. The resulting exponential couples the momenta
\( k^{>} \) and \( k^{<} \) and may therefore lead to dephasing. 

From now on, we again consider baths which are characterized by a power-law
spectrum at low frequencies, \( C(\omega )\propto \omega ^{\alpha } \) with
an exponent \( \alpha \geq 1 \). Remember that the case \( \alpha =1 \) represents
fluctuations of the magnetic flux produced by Nyquist noise of an external current
loop. For these bath spectra, it is sufficient to carry out an expansion of
the cotunneling rate to leading order in the coupling strength \( g \). The
part of the resulting expression which couples \( k^{>} \) and \( k^{<} \)
is seen to lead, after summation over all electronic states \( k^{>} \), \( k^{<} \),
\( \lambda  \), \( \bar{\lambda } \), to an {}``incoherent{}'' contribution
that washes out destructive interference but which is suppressed for low bias
voltages, as expected.

At zero temperature, the ratio of this incoherent current at \( \Phi =\Phi _{0}/2 \)
to the normal elastic cotunneling current which flows at \( \Phi =0 \) is found
to be given by the following approximate expression (up to a constant of order
$1$):

\begin{equation}
\label{ratio}
\left[ g^{2}v_{F}^{2}\int _{0}^{eV}(1-\frac{\omega }{eV})C(\omega
)d\omega \right] {1 \over {\delta \epsilon} ^2}\; .
\end{equation}

 The expression inside the brackets can be interpreted as the variance of the
fluctuating energy of a single-particle level on the ring. However, it is to
be evaluated taking into account \textit{only the fluctuations up to the frequency
corresponding to the bias voltage} and using a weight factor \( 1-\omega /eV \)
which favors low-energy transfers \( \omega  \). We have already pointed out
that the cutoff at \( eV \) is a simple consequence of energy conservation.
For a power-law bath spectrum \( C(\omega )\propto \omega ^{\alpha } \), the
integral yields a voltage dependence \( \propto V^{\alpha +1} \), so the incoherent
tunneling current goes like \( V^{\alpha +2} \). Note that \( \delta \epsilon =hv_{F}/L \)
refers to the single-particle level-spacing on the ring. The qualitative behaviour
of the cotunneling rate as a function of both external static flux and bias
voltage is shown in Fig.~(\ref{cotunnelrate}). (We remark again that the incoherent
current due to external Nyquist noise would be too small to be actually measurable
under reasonable experimental conditions, compare the discussion at the end
of section \ref{discussdephasing}).

\begin{figure}
\centerline{\hbox{  \includegraphics[width=7cm]{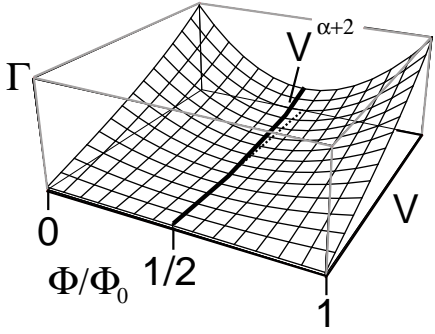} }}

\caption{\label{cotunnelrate}Schematic behavior of the cotunneling rate as a function
of static magnetic flux and bias voltage. In the ideal situation, the rate vanishes
at $\Phi_0/2$ (dotted line), while it rises as a power of the bias voltage due to the incoherent contribution resulting from the fluctuations of the flux (thick line). }
\end{figure}

Thus, we see that suppression of destructive interference does not show up in
the \textit{linear} conductance. In this sense, the fluctuations do not lead
to dephasing in the linear transport regime at zero temperature. For the Nyquist
case \( \alpha =1 \), the exponent of \( V \) is the same as that for inelastic
electronic cotunneling processes in a system with a \textit{continuum} of intermediate
electronic states \cite{averinnazarov}. Bath spectra with \( \alpha >1 \)
obviously lead to an even weaker decrease in the visibility of the interference
minimum at low bias voltages. Note that formally inserting \( \alpha =-1 \)
in \( \Gamma \propto V^{\alpha +2} \) would lead to an incoherent contribution
to the linear conductance even at \( T=0 \). However, this case is not of interest
here, since it cannot be produced by a fluctuating magnetic flux and it is not
covered by the approximations made in our calculation (in particular dropping
the \( \hat{\phi }^{2} \)-term). It would correspond to the strong force fluctuations
of an Ohmic Caldeira-Leggett bath used in the description of quantum Brownian
motion.

For bias voltages \( eV \) smaller than the single-particle energy spacing
\( \delta \epsilon  \) on the ring, dephasing is merely due to the coupling
to the fluctuating flux. At higher voltages, the inelastic cotunneling processes
become important. In these, one electron tunnels into the ring, while \textit{another}
electron goes out at the opposite electrode, thus leaving behind a particle-hole
excitation on the ring \cite{averinnazarov}. Since all the corresponding final
states are different, their contributions to the cotunneling current sum up
incoherently. Therefore, like dephasing produced by the bath, they also lead
to a nonvanishing contribution to the tunneling current at \( \Phi _{0}/2 \),
where, ideally, one should have perfect destructive interference. The number
of possibilities to create a particle-hole excitation with an energy of at most
\( eV \) is \( \propto (eV/\delta \epsilon )^{2} \), if we assume \( eV\gg \delta \epsilon  \).
In that regime, the ratio of the incoherent current contribution due to the
external bath to the electronic inelastic contribution is given by the 
bracket in (\ref{ratio}), multiplied by \( (\delta \epsilon / (E_C eV))^{2} \). The electronic inelastic
contribution will be the dominant one.

Finally, let us discuss finite temperatures. \textit{Without} the bath and as
long as \( T\ll \delta \epsilon  \), only the Fermi distributions in the electrodes
get smeared, which does not affect the tunneling current, if one takes into
account that now the tunneling processes do not only lead to an electron transport
from left to right but in the other direction as well. The presence of the bath
will introduce some temperature dependence for the incoherent current contribution
in this regime, since at finite temperatures the tunneling electron can not
only emit an energy quantum into the bath but also absorb a thermal bath excitation.
Therefore, the energy \( \omega  \) transferred to the bath now can be negative
as well. There is no restriction on the amount of energy an electron can absorb,
so there is no cutoff \( eV \) for negative \( \omega  \). At positive energy
transfers, the probability of spontaneous emission into the bath (\( \propto C(\omega ) \)
in Eq.~(\ref{ratio})) now has to be multiplied by \( n(\omega )+1 \), where
\( n(\omega ) \) is the Bose distribution function (induced emission). At negative
\( \omega  \), this is replaced by \( n(|\omega |) \), since only absorption
of \textit{thermal} excitations (not of vacuum fluctuations) is possible. 

Taking into account the thermal smearing of the electrode Fermi distributions,
the balance of left and right-going tunneling currents and the induced emission/absorption
mentioned just before, we have to perform the following replacements in (\ref{ratio}):

\begin{eqnarray}
\int _{0}^{eV}d\omega \, (1-\frac{\omega }{eV})C(\omega )\mapsto  &  & \nonumber \\
\int _{-\infty }^{+\infty }d\omega \, W(\beta ,\omega ,eV)\left( n(\left| \omega \right| )+\theta (\omega )\right) C(\left| \omega \right| )\, . &  & \label{finitetemp} 
\end{eqnarray}

Note that the factor multiplying $W$ in the integral corresponds directly to the function $P(E)$ which occurs in the theory of tunneling in a dissipative environment\cite{ingoldnazarov,gschoen}.  The function \( W \) itself represents an integral over the average of energies in
the left and right electrodes, \( \epsilon \equiv \left( \epsilon _{\lambda }+\epsilon _{\bar{\lambda }}\right) /2 \),
at a fixed energy transfer \( \omega \equiv \epsilon _{\lambda }-\epsilon _{\bar{\lambda }} \):

\begin{equation}
\label{Kfinite}
W\left( \beta ,\omega ,eV\right) \equiv \left( eV\right) ^{-1}\int d\epsilon \, f_{L}\left( 1-f_{R}\right) -f_{R}\left( 1-f_{L}\right) e^{-\beta \omega }\, .
\end{equation}

Here \( f_{L}=f_{\beta }\left( \epsilon _{\lambda }-eV/2\right)  \) and \( f_{R}=f_{\beta }\left( \epsilon _{\bar{\lambda }}+eV/2\right)  \)
are the Fermi distributions in the two electrodes. The factor of \( \exp \left( -\beta \omega \right)  \)
is due to the fact that for an electron going from right to left the energy
transferred to the bath is \( -\omega  \), so the ratio between the values
of \( n\left( \left| \omega \right| \right) +\theta \left( \omega \right)  \)
at positive and negative frequencies appears, which is just this factor. Note
that, in the low-temperature limit (\( \beta \rightarrow \infty  \)), the function
\( W \) becomes \( 1-\omega /eV \) for \( \omega <eV \) and vanishes for
\( \omega >eV \). This reproduces the left-hand side of Eq. (\ref{finitetemp})
as a special case. 

Using this formula, the incoherent tunneling current is found to be enhanced
by a temperature-dependent contribution \( \propto T^{\alpha +1}V \).

At \( T\geq \delta \epsilon  \), one would have to take into account the thermal
averaging over different electronic configurations on the ring (still at a fixed
particle number determined by charging energy and gate voltage). The perfect
destructive interference at \( \Phi _{0}/2 \) depends on the presence of an
electronic configuration which is symmetric in the occupancy of equal-energy
states having \( k>0 \) and \( k<0 \) (see discussion above). The thermal
average includes other configurations as well and therefore leads to a suppression
of the destructive interference in the \textit{elastic} tunneling current, even
without the bath. Furthermore, the electronic \textit{inelastic} contribution
is also enhanced at finite temperatures and becomes linear in the voltage \cite{averinnazarov}.

\section{Discussion}

Since we do not find dephasing in our model at zero temperature (under
the circumstances specified above), the reader may wonder how our result
is related to some apparently contradicting conclusions in the literature. 
First of all, we would like to point out that a comparison between
findings in different physical situations is not straightforward, since
there is no general definition of a ``dephasing rate''.
E.g., in recent work on an
Aharonov-Bohm ring containing a quantum dot capacitively
coupled to the Ohmic fluctuations in a gate\cite{cedraschi}, the
authors found that the coupling suppresses the magnitude of the
persistent current flowing in the ring and induces fluctuations of
this current. The effect persists even at $T=0$ and was
interpreted as dephasing at zero temperature. Formally,
the coupling assumed in the setup of Ref.~\onlinecite{cedraschi} 
is nondiagonal in the system's
eigenbasis, whereas it is diagonal in our case. This may indeed lead
to a weaker tendency towards dephasing in our model (see the
discussion at the end of Section \ref{modelsec}).  Still, we also 
find both a reduction of the persistent current (subsection
\ref{partsumsec}) and fluctuations (caused by the vector
potential fluctuations, since the momentum is
conserved), regardless of the details of the bath spectrum. 
Nevertheless, this apparently does not imply dephasing in
our interference setup (cotunneling transport situation), which has
been our main concern here and which has no analogy in the work of
Ref.~\onlinecite{cedraschi}. The transport effect depends in an
essential way on the long-time dynamics of the system and, therefore,
on the low-frequency behaviour of the bath spectrum (similarly to the
Caldeira-Leggett model or some aspects of the spin-boson model), while the reduction
of the persistent current does not. Therefore,  we have not interpreted the reduction
of the persistent current in terms of zero-temperature
dephasing, although it certainly constitutes a suppression of an
interference-related phenomenon. 
We believe that the suppression of
interference in a cotunneling setup considered here may be more
directly related to mesoscopic transport interference experiments.

Our model is certainly quite removed from the discussions about
low-temperature dephasing in weak localization. There, dephasing by
electron-electron interaction in an extended disordered system weakens
the coherent backscattering effect, which is an interference
phenomenon that is robust against both thermal and impurity averaging,
in contrast to the destructive interference considered in our
model. Since the electrons inside the metal interact all the time, it
is not obvious whether that situation is in any way analogous to the
kind of ``scattering'' situation considered here for the cotunneling
transport.


\section{Conclusions}

We have analyzed a simple model of a fluctuating magnetic flux threading an
Aharonov-Bohm ring and discussed its effects on equilibrium properties, such
as persistent current and tunneling density of states, as well as transport
properties such as the two-particle Green's function and the cotunneling current
through the ring. Particular emphasis has been put on the question of dephasing
and low-temperature behaviour. There are important qualitative differences depending
on the low-frequency behaviour of the bath power spectrum. For the case of Nyquist
noise and an arbitrary initial superposition of momentum states, an exponential
decay of off-diagonal elements of the density-matrix in time at \( T>0 \) goes
over into {}``power-law dephasing{}'' at zero temperature. However, if one
probes the coherence properties of the electronic motion on the ring by checking
for the possible suppression of destructive AB-interference in a cotunneling
setup, no such suppression is found in the \emph{linear} transport regime at
\( T=0 \). This is because the possibility for the electron to leave a trace
in the bath is diminished due to the energy conservation constraint. Our calculation
serves as an illustrative example of the difference between the {}``optics{}''
type of interference experiments (employing \emph{single} particles), in which
the semiclassical approximation and/or the usual Feynman-Vernon influence functional
may be applied to discuss dephasing, and the linear-transport interference experiments
encountered in mesoscopic physics, in which special care has to be taken in
the analysis of dephasing at low temperatures.

\section{Acknowledgments}

We would like to acknowledge valuable discussions with M.~B\"uttiker, Y.~Imry, G.-L.~Ingold,
A.~J.~Leggett, D.~Loss, Yu.~V. Nazarov, A.~Rosch, E.~Sukhorukov, A.~Zaikin and W.~Zwerger. Our
work is supported by the Swiss NSF.

\appendix

\section{Derivation of Caldeira-Leggett influence functional
using Keldysh time-ordering and Wick's theorem}
\label{Sec:influencederivation}

The influence functional for a system (variable \( \hat{x} \)) coupled linearly
to a linear bath (\( \hat{\phi } \)) is usually derived in the path-integral
picture, by {}``integrating out{}'' the bath variables \cite{1,2,weiss}.
This can be done because the bath consists of a set of uncoupled harmonic oscillators.
However, the calculation is usually quite cumbersome, although the result is
simple enough and involves only the real and imaginary parts of the bath correlator
\( \left\langle \hat{\phi }(t_{1})\hat{\phi }(t_{2})\right\rangle  \). Here
we present a derivation based on Keldysh time-ordering and Wick's theorem. The
oscillators of the bath and their action for an external driving force never
have to be considered and therefore this is probably the shortest route to the
well-known Caldeira-Leggett influence functional. Modifications of this approach
are applicable in more complicated situations as well (compare the main text).

The Feynman-Vernon influence functional is the overlap between the two bath
states which result from the action of two different (fixed) system trajectories
\( x^{>}(\cdot ) \) and \( x^{<}(\cdot ) \) onto the same initial bath state
\( \chi _{0} \). A thermal average over \( \chi _{0} \) has to be performed
at finite temperatures:

\begin{equation}
J(x^{<},x^{>})=\left\langle U(t,0|x^{<})\chi _{0}\right. \left| U(t,0|x^{>})\chi _{0}\right\rangle _{\chi _{0}}\, .
\end{equation}

Here the time-evolution operators depend on the system trajectory \( x(\cdot ) \)
via the interaction term \( \hat{V}(t)=x(t)\hat{\phi } \) in the Hamiltonian.
Using the interaction picture with respect to the bath Hamiltonian \( \hat{H}_{B} \),
we can explicitly write down \( J \) in the following form, with (anti-)time-ordering
symbols \( \hat{T} \) (\( \widetilde{\hat{T}} \)):

\begin{equation}
J=Z_{B}^{-1}tr\left[ e^{-\beta \hat{H}_{B}}\widetilde{\hat{T}}e^{+i\int _{0}^{t}x^{<}(s)\hat{\phi }(s)ds}\hat{T}e^{-i\int _{0}^{t}x^{>}(s)\hat{\phi }(s)ds}\right] \, .
\end{equation}

Here \( Z_{B}\equiv tr\exp (-\beta \hat{H}_{B}) \). Now we can use Keldysh
time-ordering to abbreviate this formally:

\begin{equation}
J=Z_{B}^{-1}tr\left[ e^{-\beta \hat{H}_{B}}\hat{T}_{K}e^{-i\oint _{K}x_{K}(s)\hat{\phi }(s)ds}\right] \, .
\end{equation}

Here \( x_{K}(s) \) is equal to \( x^{>}(s) \) (or \( x^{<}(s) \)) if \( s \)
lies on the forward (or backward) part of the Keldysh contour that runs from
\( 0 \) to \( t \) and back again. The advantage of this formal rearrangement
is that the application of Wick's theorem (leading to a linked-cluster expansion)
now becomes very simple. We immediately obtain for \( J \):

\begin{equation}
\exp \left[ -\frac{1}{2}\oint _{K}dt_{1}\oint _{K}dt_{2}\, \left\langle \hat{T}_{K}\hat{\phi }(t_{1})\hat{\phi }(t_{2})\right\rangle x_{K}(t_{1})x_{K}(t_{2})\right] \, .
\end{equation}

The brackets \( \left\langle \cdot \right\rangle  \) denote the thermal average
with respect to \( \hat{H}_{B} \). Now we can translate back the exponent by
taking into account all four possible combinations of the two times on the forward
or backward time paths: (To keep the notation short, we use subscripts for the
time arguments)

\begin{eqnarray}
& & -\frac{1}{2}\int _{0}^{t}dt_{1}\int _{0}^{t}dt_{2}\, \left\{ \left\langle \hat{T}\hat{\phi }_{1}\hat{\phi }_{2}\right\rangle x^{>}_{1}x^{>}_{2}+\left\langle \widetilde{\hat{T}}\hat{\phi }_{1}\hat{\phi }_{2}\right\rangle x^{<}_{1}x^{<}_{2}\right.  \nonumber \\
& & \left. -\left\langle \hat{\phi }_{1}\hat{\phi }_{2}\right\rangle x^{<}_{1}x^{>}_{2}-\left\langle \hat{\phi }_{2}\hat{\phi }_{1}\right\rangle x^{>}_{1}x^{<}_{2}\right\} \, .  
\end{eqnarray}

This can be simplified further by noting that the real part of all the four
different correlators appearing here is the same, since it is symmetric in the
time arguments: \( 2\Re e\left\langle \hat{\phi }_{1}\hat{\phi }_{2}\right\rangle =\left\langle \left\{ \hat{\phi }_{1},\hat{\phi }_{2}\right\} \right\rangle  \).
Therefore, the real part of the exponent is given by:

\begin{equation}
\label{Abb:influencereal}
-\frac{1}{2}\int _{0}^{t}dt_{1}\int _{0}^{t}dt_{2}\, \Re e\left\langle \hat{\phi }_{1}\hat{\phi }_{2}\right\rangle \left( x^{>}_{1}-x^{<}_{1}\right) \left( x^{>}_{2}-x^{<}_{2}\right) \, .
\end{equation}

It defines the imaginary part of the influence action \( S_{infl}[x^{>},x^{<}] \)
and describes dephasing and heating due to the fluctuations of the bath variable
\( \hat{\phi } \). 

Treating the imaginary part is only slightly more difficult. We have 

\begin{eqnarray}
\Im m\left\langle \hat{\phi }_{1}\hat{\phi }_{2}\right\rangle  & = & -\Im m\left\langle \hat{\phi }_{2}\hat{\phi }_{1}\right\rangle =\frac{1}{2i}\left\langle \left[ \hat{\phi }_{1},\hat{\phi }_{2}\right] \right\rangle  \\
\Im m\left\langle \hat{T}\hat{\phi }_{1}\hat{\phi }_{2}\right\rangle  & = & sgn(t_{1}-t_{2})\Im m\left\langle \hat{\phi }_{1}\hat{\phi }_{2}\right\rangle =-\Im m\left\langle \widetilde{\hat{T}}\hat{\phi }_{1}\hat{\phi }_{2}\right\rangle \, \nonumber.
\end{eqnarray}

In order to get rid of the \( sgn(t_{1}-t_{2}) \), we split the double time-integral
into one part where \( t_{2}<t_{1} \) and one with \( t_{2}>t_{1} \). In the
latter part, we interchange integration variables \( t_{1} \) and \( t_{2} \).
Then we obtain for the imaginary part of the exponent:

\begin{equation}
-i\int _{0}^{t}dt_{1}\int _{0}^{t_{1}}dt_{2}\, \Im m\left\langle \hat{\phi }_{1}\hat{\phi }_{2}\right\rangle \left( x^{>}_{1}-x^{<}_{1}\right) \left( x^{>}_{2}+x^{<}_{2}\right) \, .
\end{equation}

This defines the real part of the influence action and describes friction and
renormalization effects (e.g. effective mass). Note that we can bring the real
part of the exponent, (\ref{Abb:influencereal}), to a similar form by cutting
off the \( t_{2} \)-integral at \( t_{1} \) and dropping the factor \( 1/2 \)
in front of the expression, since there the integrand is symmetric in \( t_{1},t_{2} \).

In this way we have arrived at the well-known influence functional for a system
coupled linearly to a linear bath. As usual, a coupling of the form \( f(x)\hat{\phi } \)
just leads to a replacement \( x\mapsto f(x) \) in the final expression, and
something like \( \sum _{j}f_{j}(x)\hat{\phi }_{j} \) leads to a sum of the
corresponding influence actions, if the \( \hat{\phi }_{j} \) are uncorrelated.


\begin{thebibliography}{10}
\bibitem{2}A.O. Caldeira and A.J. Leggett, Ann. Phys. (N.Y.) {\bf 149}, 374
(1983); Physica {\bf 121}A, 587 (1983).
\bibitem{weiss}U. Weiss, \emph{Quantum Dissipative Systems}, (World Scientific, Singapore,
2000).
\bibitem{park}C. Park and Y. Fu, Physics Letters A {\bf 161}, 381 (1992).
\bibitem{1}R.P. Feynman and F.L. Vernon, Ann. Phys. (N.Y.) {\bf 24}, 118 (1963).
\bibitem{lossmartin}D. Loss and T. Martin, Phys. Rev. B {\bf 47}, 4619 (1993).
\bibitem{lossmullen}D. Loss and K. Mullen, Phys. Rev. B {\bf 43}, 13252 (1991).
\bibitem{sai}A. Stern, Y. Aharonov, and Y. Imry, Phys. Rev. A {\bf 41}, 3436 (1990).
\bibitem{mohanty} See P. Mohanty, E.~M.~Q. Jariwala, and R.~A. Webb,
Phys. Rev. Lett. {\bf 78}, 3366 (1997) for recent experiments,
D.~S. Golubev and A.~D. Zaikin, Phys. Rev. Lett. {\bf 81}, 1074
(1998);  {\it Physica B} {\bf 255}, 164 (1998); Phys. Rev. B {\bf 59} 9195 (1999); Phys. Rev. B {\bf 62} 14061
(2000) for the claim that there is dephasing even at $T=0$, and
I.~L. Aleiner, B.~L. Altshuler, and M.~E. Gershenson, Phys. Rev. Lett. {\bf
82}, 3190 (1999); Waves in Random Media {\bf 9}, 201 (1999) for a criticism
of this claim.
\bibitem{cohenimry}D. Cohen and Y. Imry, Phys. Rev. B {\bf 59}, 11143 (1999).
\bibitem{cedraschi}P. Cedraschi, V. V. Ponomarenko, and M. B\"uttiker, Phys. Rev. Lett. {\bf 84},
346 (2000).
\bibitem{buettrevb}P. Cedraschi and M. B\"uttiker, Phys. Rev. B {\bf 63}, 165312 (2001); Annals
of Physics (N.Y.) {\bf 289}, 1 (2001).
\bibitem{buettnato}M. B\"uttiker, cond-mat/0106149 (2001), to appear in proceedings of NATO ASI,
Geilo, Norway, April 17-27 (2001), eds. A. T. Skjeltorp and T. Vicsek (Kluwer,
Dordrecht).
\bibitem{kravtsov}V. E. Kravtsov and B. L. Altshuler, Phys. Rev. Lett. {\bf 84}, 3394 (2000).
\bibitem{mello}P. A. Mello, Y. Imry, and B. Shapiro, Phys. Rev. B {\bf 61}, 16570 (2000).
\bibitem{buettseelig}G. Seelig and M. B\"uttiker, cond-mat/0106100 (2001).
\bibitem{Ford}G. W. Ford, J. T. Lewis and R. F. O'Connell, Phys. Rev. A {\bf 37}, 4419 (1988).
\bibitem{3}V. Hakim and V. Ambegaokar, Phys. Rev. A {\bf 32}, 423 (1985).
\bibitem{schramm}P. Schramm and H. Grabert, J. Stat. Phys. {\bf 49}, 767 (1987). 
\bibitem{welton}T. A. Welton, Phys. Rev. {\bf 74}, 1157 (1948).
\bibitem{milonni}P. W. Milonni, \emph{The Quantum Vacuum} (Academic Press, San Diego, 1994).
\bibitem{averinnazarov}D. V. Averin and Yu. V. Nazarov, in: {}``Single charge tunneling{}'', ed.
H.~Grabert and M.~H.~Devoret, Plenum Press, New York (1992).
\bibitem{ingoldnazarov}G.-L. Ingold and Yu. V. Nazarov, in: ``Single charge tunneling'', loc. cit.
\bibitem{gschoen}G. Sch\"on, in: ``Quantum transport and dissipation'', T.~Dittrich et al., Wiley-VCH, Weinheim (1998).
\bibitem{keldysh}L.~V.~Keldysh, Zh. Eksp. Teor. Fiz. {\bf 47}, 1515 (1964) {[}Sov. Phys.
JETP {\bf 20}, 1018 (1965){]}. 
\bibitem{zagoskin}A. Zagoskin, \emph{Quantum Theory of Many-Body Systems}, (Springer, New York,
1998).

\end{thebibliography}
\end{document}